\documentclass[10pt, journal, twoside]{IEEEtran}

\pdfoutput=1

\usepackage{graphicx}
\usepackage{subfigure}
\usepackage{amssymb}
\usepackage{amsmath}
\usepackage{bm}

\newcounter{mytempeqncnt}

\begin{document}
%
% paper title
% can use linebreaks \\ within to get better formatting as desired
% Do not put math or special symbols in the title.
\title{An Efficient Four-Parameter Affine Motion Model for Video Coding}
%
%
% author names and IEEE memberships
% note positions of commas and nonbreaking spaces ( ~ ) LaTeX will not break
% a structure at a ~ so this keeps an author's name from being broken across
% two lines.
% use \thanks{} to gain access to the first footnote area
% a separate \thanks must be used for each paragraph as LaTeX2e's \thanks
% was not built to handle multiple paragraphs
%

\author{Li~Li,
        Houqiang~Li,~\IEEEmembership{Senior Member,~IEEE},
        Dong~Liu,~\IEEEmembership{Member,~IEEE},
        Haitao~Yang,
        Sixin~Lin,
        Huanbang~Chen,
        and Feng~Wu,~\IEEEmembership{Fellow,~IEEE}
\thanks{L. Li, H. Li, D. Liu, and F. Wu are with the CAS Key Laboratory of Technology in Geo-Spatial Information Processing and Application System, University of Science and Technology of China, Hefei 230027, China. Professor Houqiang Li is the corresponding author. (e-mail: lilimao@mail.ustc.edu.cn; lihq@ustc.edu.cn; dongeliu@ustc.edu.cn; fengwu@ustc.edu.cn)

H. Yang, S. Lin, and H. Chen are with the Media Technology Laboratory, Central Research Institute of Huawei Technologies Co., Ltd. (e-mail: haitao.yang@huawei.com; linsixin@huawei.com; chenhuanbang@huawei.com) }% <-this % stops a space
}
\markboth{submitted to IEEE Transactions on Circuits and Systems for Video Technology}%
{LI \MakeLowercase{\textit{et al.}}: An Efficient Four-Parameter Affine Motion Model for Video Coding }

% make the title area
\maketitle

% As a general rule, do not put math, special symbols or citations
% in the abstract or keywords.
\begin{abstract}
In this paper, we study a simplified affine motion model based coding framework to overcome the limitation of translational motion model and maintain low computational complexity.
The proposed framework mainly has three key contributions.
First, we propose to reduce the number of affine motion parameters from 6 to 4.
The proposed four-parameter affine motion model can not only handle most of the complex motions in natural videos but also save the bits for two parameters.
Second, to efficiently encode the affine motion parameters, we propose two motion prediction modes, i.e., advanced affine motion vector prediction combined with a gradient-based fast affine motion estimation algorithm and affine model merge, where the latter attempts to reuse the affine motion parameters (instead of the motion vectors) of neighboring blocks.
Third, we propose two fast affine motion compensation algorithms. One is the one-step sub-pixel interpolation, which reduces the computations of each interpolation.
The other is the interpolation-precision-based adaptive block size motion compensation, which performs motion compensation at the block level rather than the pixel level to reduce the interpolation times.
Our proposed techniques have been implemented based on the state-of-the-art high efficiency video coding standard, and the experimental results show that the proposed techniques altogether achieve on average 11.1\% and 19.3\% bits saving for random access and low delay configurations, respectively, on typical video sequences that have rich rotation or zooming motions.
Meanwhile, the computational complexity increases of both encoder and decoder are within an acceptable range.
\end{abstract}

% Note that keywords are not normally used for peerreview papers.

\IEEEpeerreviewmaketitle

\section{Introduction}
\label{Sec::Introduction}
Motion estimation (ME) and motion compensation (MC) are the fundamental techniques of video coding to remove the temporal redundancy between video frames.
Block matching-based ME and block-based MC have been integrated into the reference softwares of almost all the existing video coding standards, including the currently widely adopted H.264/MPEG-4 AVC \cite{Wiegand2003} and the state-of-the-art H.265/MPEG-H High Efficiency Video Coding (HEVC) \cite{Sullivan2012}.
The underlying model of block-based MC is translational motion model, which is too simple to efficiently describe the complex motions in natural videos, such as rotation and zooming.
During the development of the video coding standards, many efforts have been made to characterize the complex motions.
For example, partitioning blocks into smaller ones can handle complex motions to some extent \cite{KimKoo2012}, but may incur more overhead bits for block partitions and more motion vectors (MVs).
Therefore, since the translational motion model has not been changed, the
standard-based video coding framework is unable to represent the complex motions such as rotation and zooming efficiently.

In the year of as early as 1993, Seferidis \emph{et al.} \cite{Seferidis1993} pointed out that high-order motion models, such as affine, bilinear, and perspective motion models, were more efficient to characterize the complex motions than the translational one.
Among all the high-order motion models introduced in \cite{Seferidis1993}, affine motion model has received the most attention of research due to its simplicity.
Previous work on affine motion model can be divided into two categories: global affine motion model and local affine motion model.

For global affine motion model \cite{Wiegand2005,Li2005}, usually several groups of model parameters are used to build global affine motion models between two frames, and then each reference frame is warped several times using different model parameters to generate multiple warped references.
However, due to a limited number of global affine motion models, such methods are less capable of providing accurate motion parameters for every local motion region.
Besides, the increased number of reference frames due to the warped ones will increase the ME computations significantly.

Local affine motion model can be further categorized into mesh-based and generalized block-based.
In the mesh-based methods \cite{Nakaya1994,Huang1994}, the vertices of the mesh are known as control points, whose MVs are used to determine the motions of all the other pixels through locally variant motion models.
Since the control point is shared by the neighboring blocks, it is difficult for us
to determine the MVs of the control point through ME in a block-based rate-distortion optimization (RDO) process due to the spatial dependency between neighboring blocks.
Therefore, the mesh-based methods cannot be well integrated into the modern video coding standards.

In the generalized block-based methods \cite{Cheung2010,Huang2013}, each block can determine its own affine motion parameters.
This is consistent with the standardized video coding framework, only replacing MVs by affine motion parameters for MC.
Generalized block-based affine motion model is intuitively promising to better characterize complex motions, thereby improving coding efficiency.
However, both ME and MC under affine motion models are significantly more complex than the traditional block matching-based ME and block-based MC.
Existing works have not achieved a good balance between coding efficiency and coding complexity.
Some of them failed to achieve significantly better rate-distortion (R-D) performance, whilst others had too much complexity.

In this paper, we propose a video coding framework using the approach of generalized block-based affine motion model, which can achieve a better trade-off between coding efficiency and computational complexity.
The framework can be seamlessly integrated into the modern video coding standards, e.g., HEVC. In summary, the proposed framework mainly has the following key contributions:
\begin{itemize}
  \item A four-parameter affine motion model is studied in this paper. Different from previous works that adopted the six-parameter model, the four-parameter model has only four degrees of freedom or equivalently needs only two MVs to represent.
  It saves two parameters for each block.
  Meanwhile, this model can accurately characterize rotation, zooming, translation, and any combination of them.
  Therefore, it can handle most of the complex motions in natural videos.
  \item To efficiently encode affine motion parameters, we propose two techniques: advanced affine motion vector prediction combined with fast affine ME, and affine model merge.
  The proposed fast affine ME algorithm iteratively updates the two MVs of a block according to gradient descent.
  It was originally proposed for the six-parameter affine motion model in our previous work \cite{Li2015}, and extended for the four-parameter model herein.
  The proposed gradient-based fast affine ME algorithm can converge very fast, thus can reduce the encoding complexity significantly.
  In addition, the proposed affine model merge tries to reuse the affine motion parameters of neighboring blocks instead of regenerating a new model from the MVs of the neighbors.
  The affine model merge can make full use of the motion model correlation between neighboring blocks, and thus can improve the coding efficiency.
  \item We also propose two fast affine MC techniques to reduce both the encoding and decoding complexities.
  A one-step sub-pixel interpolation filter which can decrease the interpolation times significantly is developed to replace the previous two-step sub-pixel interpolation.
  Moreover, the block-based MC rather than the pixel-based MC is adopted for acceleration, together with an adaptive choice of block size to ensure interpolation precision.
\end{itemize}
We perform experiments to verify the efficiency of the proposed video coding framework integrated with HEVC.
Compared to HEVC main profile, our proposed techniques altogether can achieve significant bitrate savings while maintain computational efficiency.

This paper is organized as follows.
In Section \ref{Sec::related work}, we will give a brief review of the related works.
The proposed low-complexity four-parameter affine motion model based framework will be introduced in Section \ref{Sec::affine algorithm}.
The experimental results are shown in Section \ref{Sec::experimental results}.
Finally, Section \ref{Sec::conclusions} concludes this paper.

\section{Related Work}
\label{Sec::related work}
The affine motion model utilized in ME and MC can be divided into two categories: global affine motion model and local affine motion model.
For global affine motion model, Wiegand \emph{et al.} \cite{Wiegand2005} proposed to use several global affine motion models to generate several warped reference frames.
The warped reference frames were used to obtain a better prediction, and the index of the warped reference frame was needed to be transmitted to the decoder.
To reduce the overhead bitrate, Li \emph{et al.} \cite{Li2005} developed a 4-D vector quantizer to code the affine motion parameters more efficiently.
Besides, Yu \emph{et al.} \cite{Yu2009} proposed to use only one global affine motion model, and the MVs of the salient features between the original frame and reference frame were used to determine the affine motion parameters to generate a warped reference frame.
However, the global affine motion model based methods cannot provide accurate affine motion parameters for each local motion region.

The local affine motion model can be utilized in two manners: mesh-based methods and generalized block-based methods.
Nakaya and Harashima \cite{Nakaya1994} firstly proposed to use 2-D mesh to perform MC and designed a simple method to determine the MVs of the control points.
Toklu \emph{et al.} \cite{Toklu1996} proposed to add control points hierarchically to better determine the motion models of each block.
Al-Regib \emph{et al.} \cite{Al-Regib2003} further developed the method and proposed a content-based irregular mesh to better describe the object boundary.
However, due to the various block sizes in modern video coding framework, the problem of determining the MVs of control points through RDO remains difficult.

Besides the mesh-based methods, the generalized block-based methods have also been studied by many researchers.
Kordasiewicz \emph{et al.} \cite{Kordasiewicz2007} considered to derive a better prediction block through affine motion model using the surrounding translational MVs.
Besides, Cheung and Siu \cite{Cheung2010} proposed to use the neighboring information to estimate the affine motion parameters of the current block and added an affine mode into the mode decision process.
Then Narroschke and Swoboda \cite{Narroschke2013} found that the affine motion model was more suitable for the large blocks introduced in HEVC.
Huang \emph{et al.} \cite{Huang2012} extended the work in \cite{Cheung2010} for HEVC and designed the affine skip/direct mode to improve the coding efficiency.
This work was further developed to a quite complex affine MC framework and many coding modes including affine skip/direct, affine merge, and affine inter were designed to fully exploit the motion correlation between neighboring blocks \cite{Huang2013}.
Moreover, Heithausen and Vorwerk \cite{Heithausen2015} investigated and compared the performance of different kinds of high-order motion models in HEVC.
Also, with the development of the merge mode \cite{Helle2102} in HEVC, Chen \emph{et al.} \cite{Chen2015} further developed the affine skip/direct mode to incorporate with the merge mode for the translational motion model and proposed to add some temporal motion candidates into the candidate lists.
However, the previous affine merge schemes always attempted to regenerate a new affine motion model through the motion information of the neighboring blocks.
Since the neighboring blocks may correspond to different objects or have totally different motion models, the regenerated affine motion model may be inaccurate.

There is a class of local affine motion modeling algorithms designed specifically for the zooming motion in videos.
Yuan \emph{et al.}  \cite{Yuan2010} proposed to use the zooming model to generate a better motion vector predictor (MVP) for the current block.
Since this work only generated a better MVP, the R-D performance improvement was limited.
The algorithm in \cite{Yuan2012} further developed a zooming motion model to better characterize the zooming motion and proposed to use linear regression to estimate the MV of the current block from the MVs of the neighboring blocks.
Besides, Po \emph{et al.} \cite{Po2010} proposed to generate multiple zooming references using a group of model parameters and designed a sub-sampled block-matching algorithm to reduce the complexity of ME over a number of reference frames.
Kim \emph{et al.} \cite{Kim2012} proposed a 3-D diamond pattern search to reduce the number of search points during ME.

One critical issue, which hinders the adoption of affine as well as other high-order motion models, is the significant increase of ME complexity. 
In fact, in the modern video coding framework, the ME process always takes the majority of the encoding time even for translational motion model.
Due to the high complexity of ME, the fast ME algorithms \cite{Zhu2000,Zhu2002} have been hot research topics for a long time.
For example, the famous Enhanced Predictive Zonal Search (EPZS) \cite{Tourapis2002} algorithm was adopted into the H.264/AVC reference software.
HEVC reference software integrated a so-called Test Zone Search (TZS) \cite{Purnachand2012} method which was a further development of EPZS.
Both methods show quite good trade-offs between the R-D performance and encoding complexity for the ME of the translational motion model.
However, it is not easy to apply them to high-order motion models, for which more parameters need to be determined through ME.
Although there were also some algorithms trying to design fast ME algorithm for zooming motion \cite{Kim2012}, it is not easy to extend those algorithms to more general cases.
Therefore, there is an urgent need to design a fast ME algorithm for high-order motion models.

\section{The Proposed Four-parameter Affine MC Framework}
\label{Sec::affine algorithm}
The proposed four-parameter affine MC framework will be introduced from three aspects.
Firstly, we will introduce the derivation and representation of the proposed four-parameter affine motion model.
Secondly, the two methods to encode the affine MVs will be introduced in detail.
Thirdly, we will introduce the coding tools to speed up the MC process.

\subsection{The four-parameter affine motion model}
\label{subsec::basic framework}

\subsubsection{The derivation of the four-parameter affine motion model}
\label{subsubsec::derivation}
The typical six-parameter affine motion model can be described as
\begin{equation}
\label{six parameter model}
\left\{
             \begin{array}{lr}
             x' = ax + by + c &  \\
             y' = dx + ey + f &
             \end{array}
\right.
\end{equation}
where $a$, $b$, $c$, $d$, $e$, and $f$ are the six affine motion parameters.
The $(x,y)$ and $(x',y')$ are the coordinates of the same pixel before and after the transform of the affine motion model.
In essence, an affine transform is any transform that preserves lines and parallelism.
Therefore, the affine motion model can characterize translation, rotation, zooming, shear mapping, and so on.
However, the most common motions in daily videos only include six kinds of typical camera motions (camera track, boom, pan, tilt, zoom, and roll) and the typical object motions (translation, rotation, and zooming), for which six model parameters are more than necessary.
It should be noted that the rotation here means the object rotates in a 2-D plane that is parallel with the camera.
Also, the object zooming can be characterized using an affine motion model only if the relative distance between the object and camera keeps unchanged or the object has a planar surface.
Since this paper focuses on the local affine model, we can assume that a local block has a planar surface as long as the block is small enough.
In the following, the typical object motions will be used as examples to explain the physical interpretation of the proposed four-parameter affine motion model.

\begin{figure}[tb]
\centering
\subfigure[ rotation + translation ]
{
\includegraphics[width=0.2\textwidth]{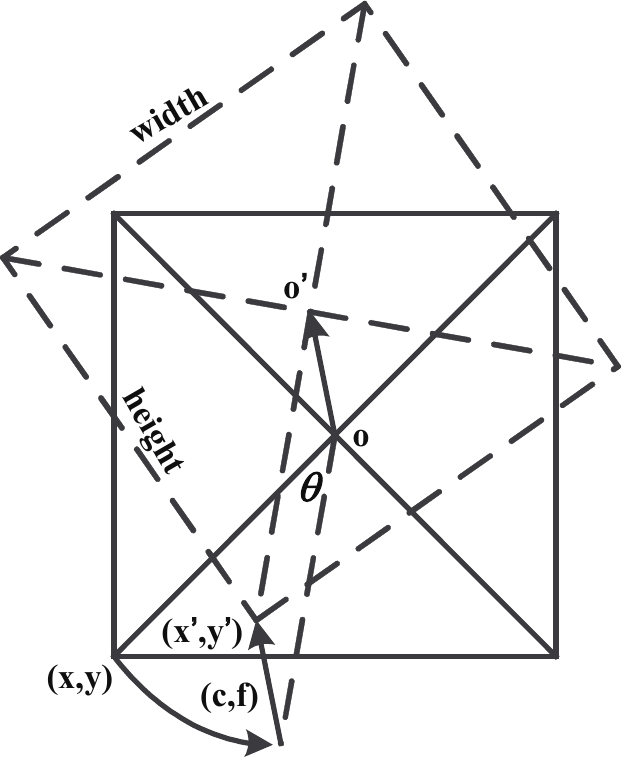}
}
\subfigure[ zooming + translation ]
{
\includegraphics[width=0.18\textwidth]{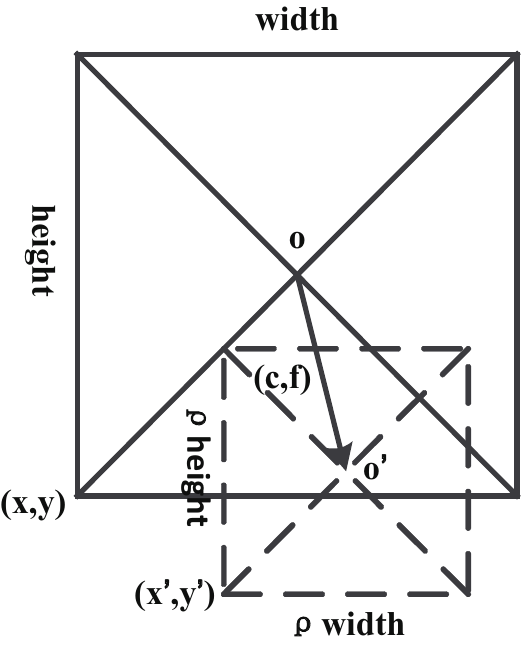}
}
\subfigure[ rotation + zooming + translation ]
{
\includegraphics[width=0.21\textwidth]{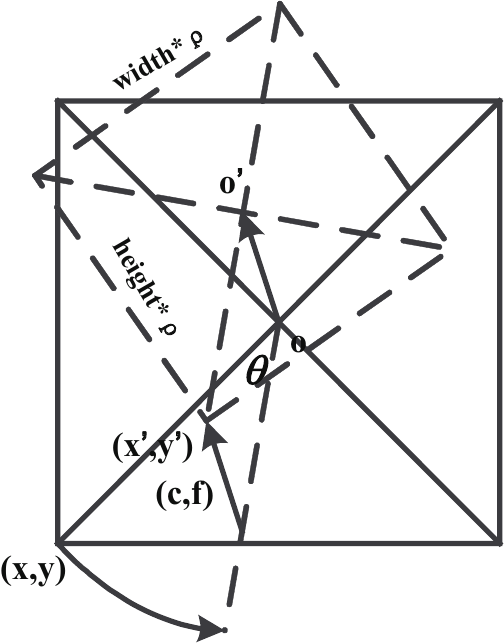}
}
\caption{Four-parameter affine model}
\label{four-parameter affine model}
\end{figure}

In fact, as shown in Fig.\ref{four-parameter affine model} (a), if only the combination of rotation and translation is needed to be characterized, the relationship between the coordinates of the same pixel before and after the transformation can be described as
\begin{equation}
\label{rotation model}
\left\{
             \begin{array}{lr}
             x' = \cos{\theta} \cdot x + \sin{\theta} \cdot y + c &  \\
             y' = -\sin{\theta} \cdot x + \cos{\theta} \cdot y + f &
             \end{array}
\right.
\end{equation}
where $\theta$ is the rotation angle.
Besides, as shown in Fig. \ref{four-parameter affine model} (b), if only the combination of zooming and translation is to be characterized, the relationship can be described as
\begin{equation}
\label{zooming model}
\left\{
             \begin{array}{lr}
             x' = \rho \cdot x + c &  \\
             y' = \rho \cdot y + f &
             \end{array}
\right.
\end{equation}
where $\rho$ is the zooming factor in both $x$ and $y$ directions, respectively.

Both the combinations of rotation/zooming and translation need three parameters to characterize.
If we combine the rotation, zooming, and translation together, four parameters will be needed and the relationship can be described as
\begin{equation}
\label{rotation and zooming model}
\left\{
             \begin{array}{lr}
             x' = \rho \cos{\theta} \cdot x + \rho \sin{\theta} \cdot y + c &  \\
             y' = -\rho \sin{\theta} \cdot x + \rho \cos{\theta} \cdot y + f &
             \end{array}
\right.
\end{equation}
If we replace $\rho \cos{\theta}$ and $\rho \sin{\theta}$ with $(1+a)$ and $b$, (\ref{rotation and zooming model}) can be rewritten as
\begin{equation}
\label{four parameter model}
\left\{
             \begin{array}{lr}
             MV_{(x,y)}^h = x' - x = ax + by + c &  \\
             MV_{(x,y)}^v = y' - y = -bx + ay + f &
             \end{array}
\right.
\end{equation}
where $MV_{(x,y)}^h$ and $MV_{(x,y)}^v$ are the horizontal and vertical components of MV for the position $(x,y)$.
Eq. (\ref{four parameter model}) is the four-parameter affine motion model used in this paper to accurately characterize the combination of rotation, zooming, and translation.

Comparing the six-parameter affine motion model in (\ref{six parameter model}) with the four-parameter affine motion model in (\ref{four parameter model}), it can be obviously seen that fewer parameters will be calculated under the four-parameter affine motion model for each block thus the decoding complexity can be slightly reduced. 
Besides, the encoding complexity will also be reduced since the proposed fast affine ME algorithm which will be introduced later on can converge faster under the four-parameter affine motion model. 
Last but not least, the four-parameter affine motion model can lead to better R-D performance for most natural sequences due to the bits savings of header information.
\vspace{3ex}

\subsubsection{The representation of the four-parameter affine motion model}

\begin{figure}[tb]
\centering
\includegraphics[width=0.25\textwidth]{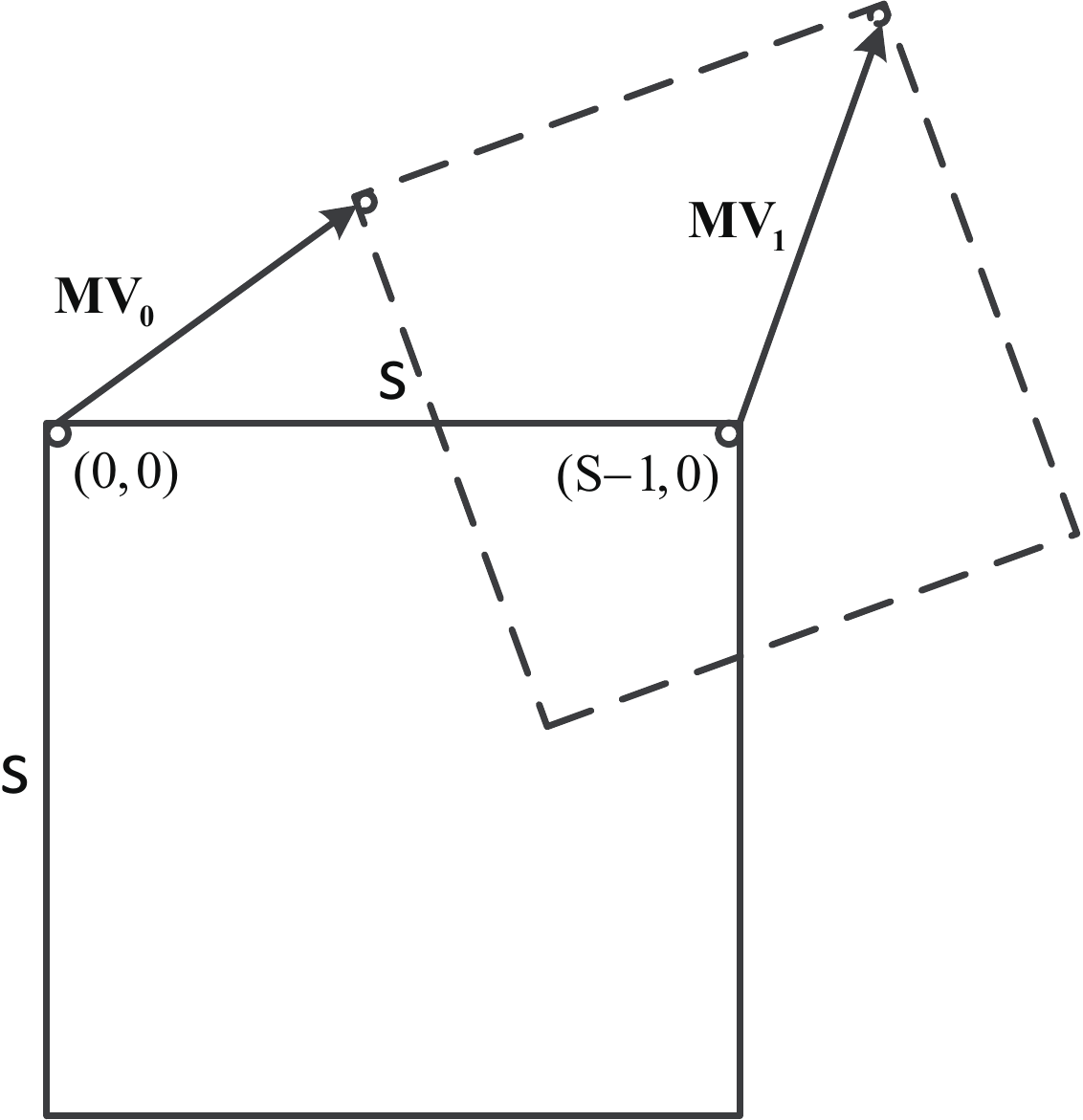}
\caption{Affine motion model representation}
\label{affine_model}
\end{figure}

According to (\ref{four parameter model}), there are four unknown model parameters.
Instead of these four parameters, we can also use two MVs to equivalently represent the model because using MVs is more consistent with existing video coding framework.
Those two MVs can be chosen at any locations of the current block for representing the motion model.
In this paper, we choose the MVs at the top left and top right locations of the current block, because these two locations are adjacent to the previously reconstructed blocks, and the corresponding MVs can be more accurately predicted.
In a typical $S \times S$ block as shown in Fig. \ref{affine_model}, if we denote the MV of top left pixel $(0,0)$ as $\bm{MV_0}$ and the MV of top right pixel $(S-1,0)$ as $\bm{MV_1}$, the four unknown model parameters $a$, $b$, $c$, and $f$ can be solved as follows according to (\ref{four parameter model}).
\begin{equation}
\label{eq:solution}
\left\{
             \begin{array}{lr}
             a = \frac{MV_1^h-MV_0^h}{S-1}~~~~~c = MV_0^h     \\
             b = -\frac{MV_1^v-MV_0^v}{S-1}~~~~f = MV_0^v     
             \end{array}
\right.
\end{equation}
Then (\ref{four parameter model}) can be expressed as a linear combination of $\bm{MV_0}$ and $\bm{MV_1}$,
\begin{equation}
\label{linear model}
\left\{
             \begin{array}{lr}
             MV_{(x,y)}^h = \sum_{k=0}^{1}m_k MV_k^h + \sum_{k=0}^{1}n_k MV_k^v &  \\
             MV_{(x,y)}^v = - \sum_{k=0}^{1}n_k MV_k^h + \sum_{k=0}^{1}m_k MV_k^v &
             \end{array}
\right.
\end{equation}
where $m_0$, $m_1$, $n_0$, and $n_1$ are equal to $(1-\frac{x}{S-1})$, $\frac{x}{S-1}$, $\frac{y}{S-1}$, and $-\frac{y}{S-1}$, respectively.
$MV_k^h$ and $MV_k^v$ are the horizontal and vertical parts of $\bm{MV_k}$.
It should be noted that $m_0$, $m_1$, $n_0$, and $n_1$ are all related to the coordinate of the current pixel.
Eq. (\ref{linear model}) can also be written in a vector form,
\begin{equation}
\label{vector form}
\bm{MV(p)} = \bm{A(p)} \cdot \bm{MV_c}^T
\end{equation}
where $\bm{p} = (x,y)$,
\begin{equation}
\label{AP}
\bm{A(p)} = \begin{bmatrix} m_0 & m_1 & n_0 & n_1 \\ -n_0 & -n_1 & m_0 & m_1 \end{bmatrix}
\end{equation}
\begin{equation}
\label{MVC}
\bm{MV_c} = [MV_0^h, MV_1^h, MV_0^v, MV_1^v]
\end{equation}
Eq. (\ref{vector form}) shows that $\bm{MV_0}$ and $\bm{MV_1}$ control the motions of all the pixels in a block.
If we know the motions of all the pixels in the block, then the MC process can be performed and the corresponding prediction block can be obtained.
Therefore, the key problem becomes how to determine $\bm{MV_0}$ and $\bm{MV_1}$.
In this paper, the precisions of both $\bm{MV_0}$ and $\bm{MV_1}$ are set as $\frac{1}{4}$ pixel to get a good trade-off between the affine motion model accuracy and overhead bits.
This is also consistent with HEVC.

\subsection{Affine motion estimation}
\label{subsec::affine MV coding}
There are usually two methods to determine the translational MV in a typical encoder of HEVC (HM or x265): AMVP mode combined with a fast ME algorithm and merge mode. 
The AMVP mode constructs an MVP candidate list for the translational MV and the ME process is used to get the optimal MV for MC.
The merge mode constructs a merge candidate list and reuses the motion information of the neighboring blocks.
Analogously, we also design two methods in this paper to determine the affine MVs: advanced affine motion vector prediction (AAMVP) mode combined with a fast affine ME method and affine model merge (AMM) mode.

\vspace{3ex}

\subsubsection{AAMVP}
\label{subsubSec::AAMVP}

\begin{figure}[tb]
\centering
\includegraphics[width=0.25\textwidth]{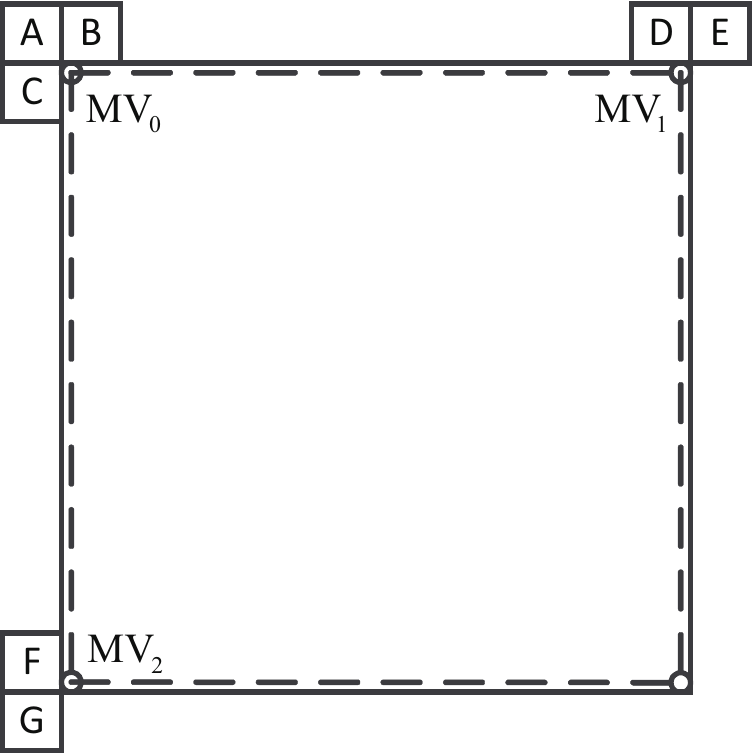}
\caption{Advanced affine motion vector prediction}
\label{Affine MVP}
\end{figure}

Similar to the AMVP mode, the AAMVP mode tries to obtain a candidate list of MV tuples to predict $(\bm{MV_0}, \bm{MV_1})$.
The construction of AAMVP candidate list is performed in three steps.
Firstly, we find the available MVP candidates for $\bm{MV_0}$, $\bm{MV_1}$, and $\bm{MV_2}$ (the MV of the bottom left corner) separately.
As shown in Fig. \ref{Affine MVP}, the MVs of neighboring blocks A, B, and C are used as the candidates for the MVP of $\bm{MV_0}$, the MVs of neighboring blocks D and E are used for the MVP of $\bm{MV_1}$, and the MVs of neighboring blocks F and G are used for the MVP of $\bm{MV_2}$.
We will use the derivation of the candidates for $\bm{MVP_0}$ (the MVP of $\bm{MV_0}$) as an example to explain.
The availability of the MVs of blocks A, B, and C will firstly be checked (for example, intra mode means unavailable).
If available, we will then check whether the MVs of blocks A, B, and C are pointing to the same reference frame as the given one.
If yes, the candidates for $\bm{MVP_0}$ are found.
If not, scaling operations are applied to make the MVs of blocks A, B, and C point to the same reference frame as the given one so as to obtain the candidates for $\bm{MVP_0}$.
The derivation processes of $\bm{MVP_1}$ and $\bm{MVP_2}$ (the MVP of $\bm{MV_1}$ and $\bm{MV_2}$) are similar to that of $\bm{MVP_0}$.

Secondly, a candidate list of MV tuples is constructed.
The $\bm{MVP_0}$ and $\bm{MVP_1}$ are combined to get a candidate list.
There are two constraints for $\bm{MVP_0}$ and $\bm{MVP_1}$.
On one hand, the $\bm{MVP_0}$ and $\bm{MVP_1}$ should not be equal since the equal $\bm{MVP_0}$ and $\bm{MVP_1}$ means translational motion model.
On the other hand, the differences between $\bm{MVP_0}$ and $\bm{MVP_1}$ in both horizontal and vertical directions should not be larger than a predefined threshold.
Too large differences mean $\bm{MVP_0}$ and $\bm{MVP_1}$ are probably from different objects, which makes the combination of $\bm{MVP_0}$ and $\bm{MVP_1}$ in a single motion model unreasonable.
The threshold is set as half of the block size in our implementation.
Since we may find multiple candidates through the above steps, they should be put into the candidate list in a specified order.
Note that, we have the following relationships between $\bm{MV_0}$, $\bm{MV_1}$, and $\bm{MV_2}$ (refer to Fig. \ref{Affine MVP} and Eq. (\ref{four parameter model})).
\begin{equation}
\label{eq:relationship}
\left\{
             \begin{array}{lr}
             MV_1^h = a \cdot (S-1) + MV_0^h          \\
             MV_2^h = b \cdot (S-1) + MV_0^h          \\
             MV_1^v = -b \cdot (S-1) + MV_0^v         \\
             MV_2^v = a \cdot (S-1) + MV_0^v
             \end{array} 
\right.
\end{equation}
These relationships can be easily converted to the following constraints for $\bm{MV_0}$, $\bm{MV_1}$, and $\bm{MV_2}$.
\begin{equation}
\label{eq:obvervation}
\left\{
             \begin{array}{lr}
             MV_1^h - MV_0^h = MV_2^v - MV_0^v         \\
             MV_0^v - MV_1^v = MV_2^h - MV_0^h
             \end{array} 
\right.
\end{equation}
Since the better MVP tuple will be the one more approximated to the MV tuple, we calculate a criterion named $DMV$ as follows.
\begin{equation}
\label{Difference of MV}
\begin{split}
DMV & = | (MVP_{1}^h - MVP_{0}^h) - (MVP_{2}^v - MVP_{0}^v) | \\
   & + | (MVP_{0}^v - MVP_{1}^v) - (MVP_{2}^h - MVP_{0}^h) |
\end{split}
\end{equation}
The smaller $DMV$ value means that the combination of $\bm{MVP_0}$ and $\bm{MVP_1}$ is more probable to form a real affine motion model and therefore, it should be put in the relatively earlier position in the candidate list.

Thirdly, if the number of MV tuples in the candidate list is less than the maximum number of candidates, each component of the MV tuple set as the translational motion is added to the candidate list to guarantee parsing robustness \cite{Li2011}.
The maximum number of MV tuples is set as $2$ to be consistent with HEVC AMVP mode.
After the above steps, we will have a candidate list of MV tuples with two candidates.

\vspace{3ex}

\subsubsection{Fast affine ME}
\label{Fast Affine ME}
After the derivation of MVP tuple candidate lists, the affine ME needs to be performed to find the optimal parameters, i.e. two MVs for a block.
A straightforward method is to search all the combinations of $\bm{MV_0}$ and $\bm{MV_1}$ within a predefined search range $R$.
However, such a method will lead to the complexity of $O(R^4)$ since two MVs are needed to be jointly determined. 
Huang et al. \cite{Huang2013} provide a simplified fast ME method to optimize one MV out of the two MVs iteratively.
In this case, the complexity will be reduced to $O(R^2)$ as the two MVs are optimized independently.
However, optimizing one MV out of the two MVs iteratively may be unable to achieve the optimal R-D performance.
Moreover, since the affine MC is rather complex, the above mentioned ME algorithms are unable to achieve acceptable encoding complexity.
Therefore, we propose a gradient-based fast affine ME algorithm which can solve the two MVs simultaneously at each iteration and converge to the optimal combination quickly.
The encoding complexity will be determined by the iteration times in the proposed algorithm.
And according to our empirical study, $6$ and $8$ times of iteration will be enough for the uni-directional and bi-directional prediction, respectively.
Therefore, the proposed algorithm can reduce the encoder complexity significantly compared with previously studied affine ME algorithms.

The essence of the fast affine ME algorithm is to adjust $\bm{MV_0}$ and $\bm{MV_1}$ to minimize the mean square error (MSE) between the current block and the prediction block.
The start search point of affine ME is the best MV tuple among MV tuples in the AAMVP candidate list and the MV tuple with each component equal to the best translational motion.
The MSE between the current block and the prediction block can be expressed as
\begin{equation}
\label{MSE}
MSE = \sum\limits_{\bm{p}\in{\bm{B}}}(Pic_{org}(\bm{p})-Pic_{ref}(\bm{p}+\bm{MV(p)}))^2
\end{equation}
where $\bm{B}$ is a collection of all the pixels in the current block, and $\bm{p}$ is the position of the current pixel in the current picture.
$Pic_{org}$ is the current picture. $Pic_{ref}$ is the reference picture.
$\bm{MV(p)}$ is the MV of position $\bm{p}$.

Define that at the $i^{th}$ iteration, the MV of position $\bm{p}$ is $\bm{MV^i(p)}$.
Assume that the MVs in the corner positions $\bm{MV^i_c}$ will change by $\bm{dMV^i_c}$ to obtain the minimum MSE between the current block and the prediction block in the next iteration, then according to (\ref{vector form}), the change of MVs for all the pixels in the block can be expressed as
\begin{equation}
\label{MVPixel}
\begin{split}
\bm{MV^{i+1}(p)} & = \bm{A(p)} \cdot (\bm{(MV^i_c)^T} + \bm{(dMV^i_c)^T})\\
                 & = \bm{MV^i(p)} + \bm{A(p)} \cdot \bm{(dMV^i_c)^T}
\end{split}
\end{equation}
Then $Pic_{ref}(\bm{p}+\bm{MV^{i+1}(p)})$ can be calculated through
\begin{equation}
\label{picRefExpress}
\begin{split}
   & Pic_{ref}(\bm{p}+\bm{MV^{i+1}(p)})     \\
 = & Pic_{ref}(\bm{p} + \bm{MV^i(p)} + \bm{A(p)} \cdot \bm{(dMV^i_c)^T} )      \\
 = & Pic_{ref}(\bm{q} + \bm{A(p)} \cdot \bm{(dMV^i_c)^T} )
\end{split}
\end{equation}
where $\bm{q}$ is the corresponding position of $\bm{p}$ in the reference block in the $i^{th}$ iteration.
Using the Taylor's expansion and ignoring the high-order terms, we have
\begin{equation}
\label{picRefTaylor}
\begin{split}
   & Pic_{ref}(\bm{p}+\bm{MV^{i+1}(p)})     \\
 = & Pic_{ref}{\bm{(q)}} + \bm{Pic'_{ref}(q)} \cdot \bm{A(p)} \cdot \bm{(dMV^i_c)^T}
\end{split}
\end{equation}

As mentioned above, the optimization target is to select the best $\bm{dMV^i_c}$ by minimizing the MSE,
\begin{equation}
\label{optimization target}
\min\limits_{\bm{dMV^i_c}} \sum\limits_{\bm{p}\in{B}}(Pic_{org}(\bm{p})-Pic_{ref}(\bm{p}+\bm{MV^{i+1}(p)}))^2
\end{equation}
Combining (\ref{picRefTaylor}) and (\ref{optimization target}), we will have
\begin{equation}
\label{optimization target final}
\min\limits_{\bm{dMV^i_c}} \sum\limits_{\bm{p}\in{B}}(e(\bm{p})-\bm{Pic_{ref}'(q)} \cdot \bm{A(p)} \cdot \bm{(dMV^i_c)^T})^2
\end{equation}
where $e(\bm{p})$ is equal to $(Pic_{org}(\bm{p})-Pic_{ref}(\bm{q}))$.
Formula (\ref{optimization target final}) is actually an unconstrained optimization problem.
By setting to zero the gradients with respect to $\bm{dMV^i_c}$, we can obtain
\begin{equation}
\label{equationToSolve}
\begin{split}
& \sum\limits_{\bm{p}\in{B}}\bm{Pic_{ref}'(q)}\bm{A(p)}_l \bm{Pic_{ref}'(q)} \bm{A(p)} \bm{(dMV_c^i)^T}  \\
= &  \sum\limits_{\bm{p}\in{B}} e(\bm{p}) \bm{Pic_{ref}'(q)} \bm{A(p)}_l  ~~~~~~~~  l = 1, 2, 3, 4
\end{split}
\end{equation}
where $\bm{A(p)}_l$ represents the $l^{th}$ column of matrix $\bm{A(p)}$.
Formula (\ref{equationToSolve}) is actually a system of linear equations.
$e(\bm{p})$ can be calculated after the $i^{th}$ iteration by subtracting the prediction block from the original block.
Both $\bm{A(p)}$ and $\bm{A(p)}_l$ are known values according to (\ref{vector form}).
$\bm{Pic_{ref}'(q)}$ is the gradient value at pixel $\bm{q}$ in the reference picture, which can be estimated using the Sobel operator as shown in Eq. (\ref{sobel}).
\begin{figure*}[tb]
\normalsize
\setcounter{mytempeqncnt}{\value{equation}}
\small
\begin{equation}
\label{sobel}
\bm{Pic_{ref}'(x,y)}=
\left[
             \begin{array}{lcl}
             ( (Pic_{ref}(x+1,y+1) - Pic_{ref}(x-1,y+1))               &   &    (Pic_{ref}(x+1,y+1) - Pic_{ref}(x+1,y-1))          \\
               + (Pic_{ref}(x+1,y-1) - Pic_{ref}(x-1,y-1))             & , &  + (Pic_{ref}(x-1,y+1) - Pic_{ref}(x-1,y-1))         \\
               +  2 \times (Pic_{ref}(x+1,y) - Pic_{ref}(x-1,y))) / 8  &   &  + 2 \times (Pic_{ref}(x,y+1) - Pic_{ref}(x,y-1))) / 8
             \end{array} 
\right]
\end{equation}
\end{figure*}

Therefore, for each iteration, just a simple system of linear equations needs to be solved to get the $\bm{dMV_c}$.
If all components of $\bm{dMV_c^i}$ are $0$ after the $i$th iteration, the $\bm{MV_c^i}$ will be used to get the prediction block. 
Different from the traditional fast ME algorithms which can only find the best MV one by one, both $\bm{MV_0}$ and $\bm{MV_1}$ can be found out simultaneously through the proposed gradient-based fast affine ME algorithm.
Therefore, the proposed algorithm can simultaneously guarantee the R-D performance and reduce the encoder complexity significantly.

After the two affine MVs are determined, the affine MVP tuple in the affine MVP tuple list which will lead to smaller affine MVD will be used as the final affine MVP tuple and the two corresponding affine MVDs will be encoded in a similar way as the MVD for the translational motion model. 
Such a scheme will lead to about two times of bits cost per prediction unit (PU) since two MVDs are transmitted per PU. 
However, since the affine motion model can improve the prediction accuracy, the number of PUs will reduce significantly due to the use of large blocks, which will lead to less number of MVDs and fewer bits for header information.
Besides, the residue bits will also decrease obviously due to the improved prediction precision brought by the affine motion model.

\vspace{3ex}

\subsubsection{AMM}
\label{subsubSec::AMM}

\begin{figure}[tb]
\centering
\includegraphics[width=0.35\textwidth]{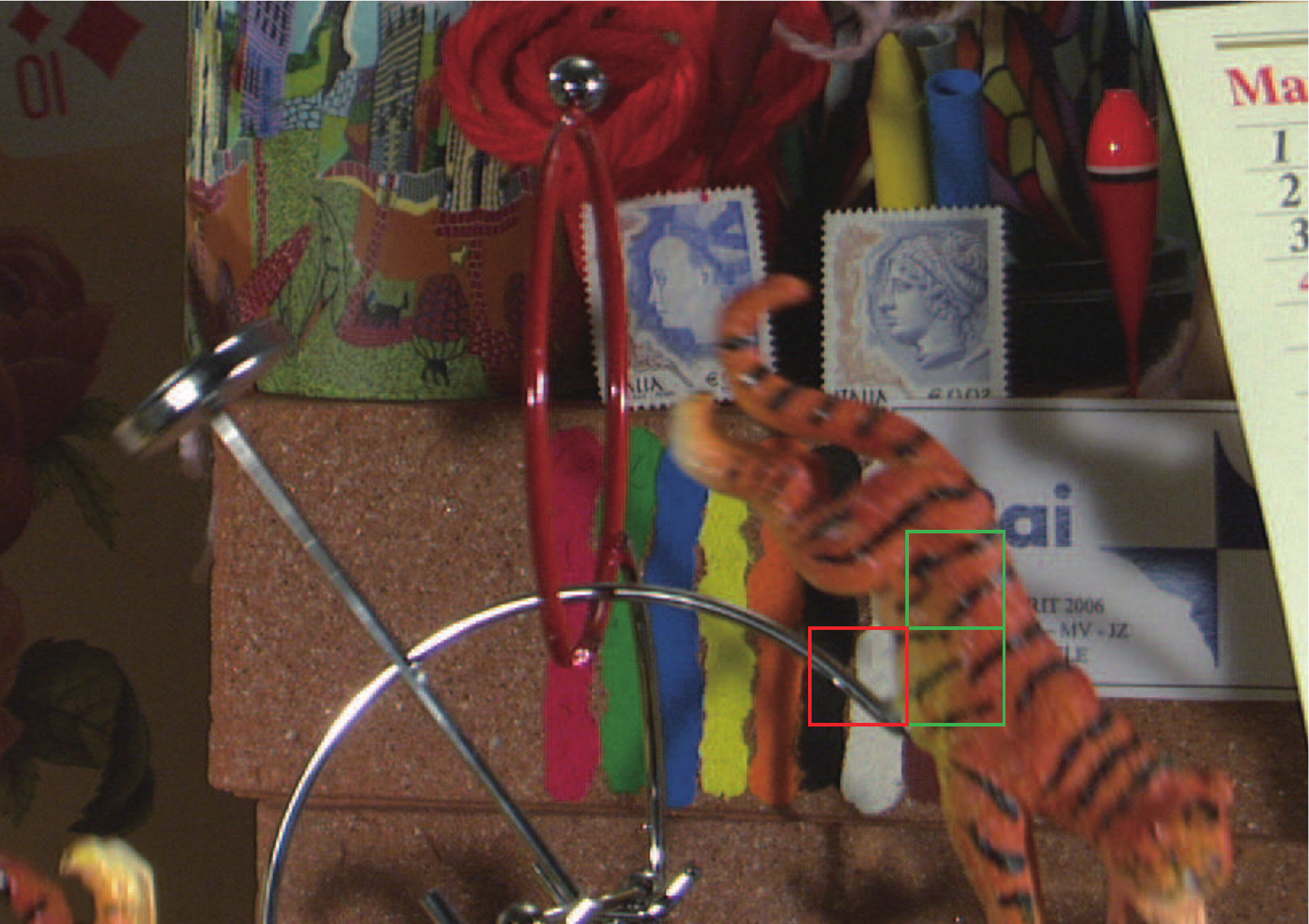}
\caption{Affine model merge example}
\label{AMM example}
\end{figure}

\begin{figure}[tb]
\centering
\includegraphics[width=0.35\textwidth]{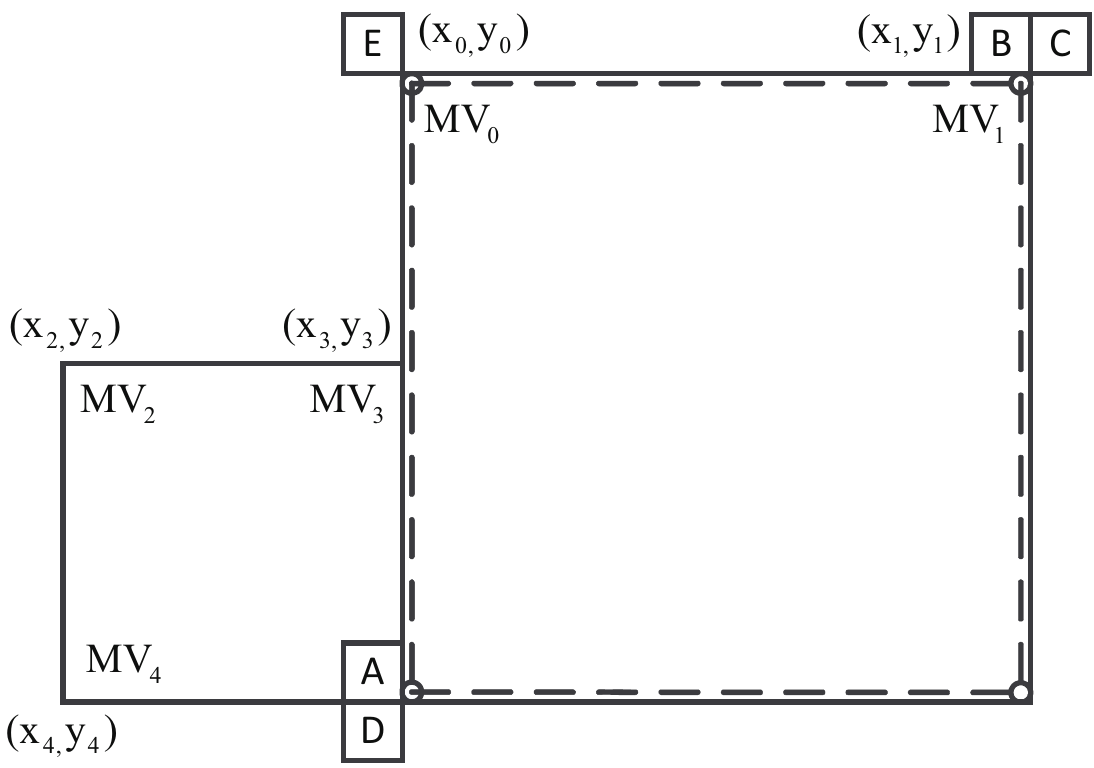}
\caption{Affine model merge candidate position}
\label{Affine model merge}
\end{figure}

Different from the existing affine merge mode which tries to regenerate an affine motion model according to the neighboring motion information \cite{Huang2012,Chen2015}, the AMM mode fully reuses the affine motion model of the neighboring blocks that also use affine mode (including AAMVP and AMM mode).
It should be emphasized that the AMM mode is used only when at least one of the neighboring blocks uses affine mode.
Fig. \ref{AMM example} gives a typical example showing the difference between the AMM and existing affine merge mode.
In Fig. \ref{AMM example}, the two green squares represent two neighboring CTUs.
In this case, the two CTUs are within the same object that is rotating.
Thus, it implies that the two CTUs probably share the same affine motion model parameter $\theta$.
The situation is similar for zooming motion and the zooming factor $\rho$ is the same for neighboring CTUs.
Therefore, the reuse of the affine motion model of the neighboring blocks means we can reuse the parameters $a$ and $b$ in (\ref{four parameter model}) since the zooming factor and rotation angle are the same for the neighboring blocks within one object.
Note however that the parameters c and f may be different for neighboring blocks.
This is indeed the reuse of model parameters, rather than using the MVs of neighboring blocks, as previous work did \cite{Chen2015}.
In the previous work, the regenerated model from the neighboring red and green blocks may lead to inaccurate model parameters.

To reuse the affine motion model of the neighboring blocks, we should firstly traverse the neighboring blocks to find the blocks using affine mode.
The search order of the AMM candidates is A, B, C, D, and E as shown in Fig. \ref{Affine model merge}, which is the same as the merge mode for translational motion model in HEVC.
If A, B, C, D, or E uses affine motion model, the candidate is added to the AMM candidate lists.
If no neighboring blocks use affine motion model, the AMM mode will be skipped for the current block.
The number of AMM candidates is set as 1 to reduce the header bits for AMM index.

Then we will use the neighboring affine motion parameters to derive the affine motion parameters of the current block.
As shown in Fig. \ref{Affine model merge}, since the top left pixel $(x_0, y_0)$ with motion $\bm{MV_0}$ and top right pixel $(x_1, y_1)$ with motion $\bm{MV_1}$ determine the affine motion parameters of the current block, we will deduce the $\bm{MV_0}$ and $\bm{MV_1}$ using the rule of the same $a$ and $b$ in neighboring blocks.
In the following, block A will be used as an example to explain the detailed deduction process.
Firstly, we will find the PU containing the block A and obtain the motion information of the PU: the top left pixel $(x_2, y_2)$ with motion $\bm{MV_2}$, the top right pixel $(x_3, y_3)$ with motion $\bm{MV_3}$, the bottom left pixel $(x_4, y_4)$ with motion $\bm{MV_4}$.
It should be noted that if block A uses affine mode, it means that $\bm{MV_2}$, $\bm{MV_3}$, and $\bm{MV_4}$ are with the same inter direction (forward, backward, or bi-direction) and reference index.
Then we can calculate the $\bm{MV_0}$ of the current block according to the relative position of the current position with the neighboring PU,
\begin{equation}
\label{MV0}
\bm{MV_0} = \bm{MV_3} + \frac{(y_3 - y_0) \times (\bm{MV_4} - \bm{MV_2})}{(y_4 - y_2)}
\end{equation}
Then the $\bm{MV_1}$ of the current block can be calculated using the rule of the same $a$ and $b$ in the neighboring blocks,
\begin{equation}
\label{MV1}
\bm{MV_1} = \bm{MV_0} + \frac{(x_1 - x_0) \times (\bm{MV_3} - \bm{MV_2})}{(x_3 - x_2)}
\end{equation}
The $\bm{MV_0}$ and $\bm{MV_1}$ can be calculated in a similar way if the block B, C, D, or E uses affine mode.
The AMM mode with residue and AMM skip mode without residue are both supported in our scheme.

\subsection{Fast coding tools for Affine MC}
\label{subsec::fast ME and MC}
The complexity of affine MC mainly comes from two aspects: the times of interpolation and the complexity of each interpolation.
We have mainly designed two coding tools focusing on these two aspects to speed up the affine MC process.
The first one is to design a one-step sub-pixel interpolation filter to decrease the complexity of each interpolation.
The second one is to use the affine interpolation-precision-based adaptive block size MC instead of pixel-based MC to decrease the times of interpolation.

\vspace{3ex}

\subsubsection{One-step sub-pixel interpolation filter}
\label{subsubsec::one-step IF}
To obtain the affine prediction block with non-integer MVs, for the Luma component, the traditional two-step interpolation filter \cite{Huang2013} will first interpolate the $1/4$ pixel accuracy using Discrete Cosine Transform based Interpolation Filter (DCTIF), which is the interpolation filter for translational MC in HEVC.
Then the bilinear interpolation will be performed if the MV is beyond $1/4$ pixel accuracy.
The situation is similar for Chroma component, the $1/8$ pixel accuracy is firstly interpolated using DCTIF and then the bilinear interpolation is applied for higher pixel accuracy.

The traditional two-step interpolation method mainly has three shortcomings.
Firstly, the computational complexity of the traditional method is much higher compared with the translational MC.
To interpolate a pixel higher than $1/4$ pixel accuracy, up to four DCTIF and one bilinear interpolation operations should be performed, which will bring quite significant complexity burdens to both the encoder and decoder.
Secondly, the bilinear interpolation filter is unable to achieve a satisfactory R-D performance for the fractional interpolation \cite{Lv2012}.
Thirdly, the arbitrary MV precision is unfriendly to the hardware implementation.

To overcome the disadvantages brought by the traditional interpolation filter, we will first determine the MV limitation precision to prevent the unfriendly arbitrary MV precision.
The MV limitation precisions of Luma and Chroma components are set as $1/64$ to obtain a balance between R-D performance and hardware implementation.
According to our empirical study, higher interpolation precision beyond $1/64$ brings out only little improvement in MC accuracy.
Besides, a one-step sub-pixel interpolation filter designed based on the principle of DCTIF is used to interpolate the $1/64$ pixel precision.
Since the DCTIF outperforms the bilinear interpolation filter in the aspect of interpolating the fractional pixels \cite{Lv2012}, the proposed one-step sub-pixel interpolation filter can achieve better R-D performance compared with the traditional two-step interpolation filter.
This will also be verified by the experimental results shown in the next section. 
Moreover, the one-step sub-pixel interpolation filter can obtain the prediction pixel using only one DCTIF operation for all the pixel precisions and thus can reduce the MC complexity significantly.
To unify with the translational interpolation filter in HEVC, the taps of interpolation filter for Luma and Chroma components are set as $8$ and $4$, respectively.
Due to the limited space, the interpolation filter coefficients for Luma and Chroma components are not shown in this paper.
More detailed interpolation coefficients can be found in \cite{Lin2015}.

\vspace{3ex}

\subsubsection{Affine interpolation-precision-based MC}
\label{subsubsec::block MC}

\begin{figure}[tb]
\centering
\subfigure[ pixel-based MC, $9 \times 16$ interpolations ]
{
\includegraphics[width=0.3\textwidth]{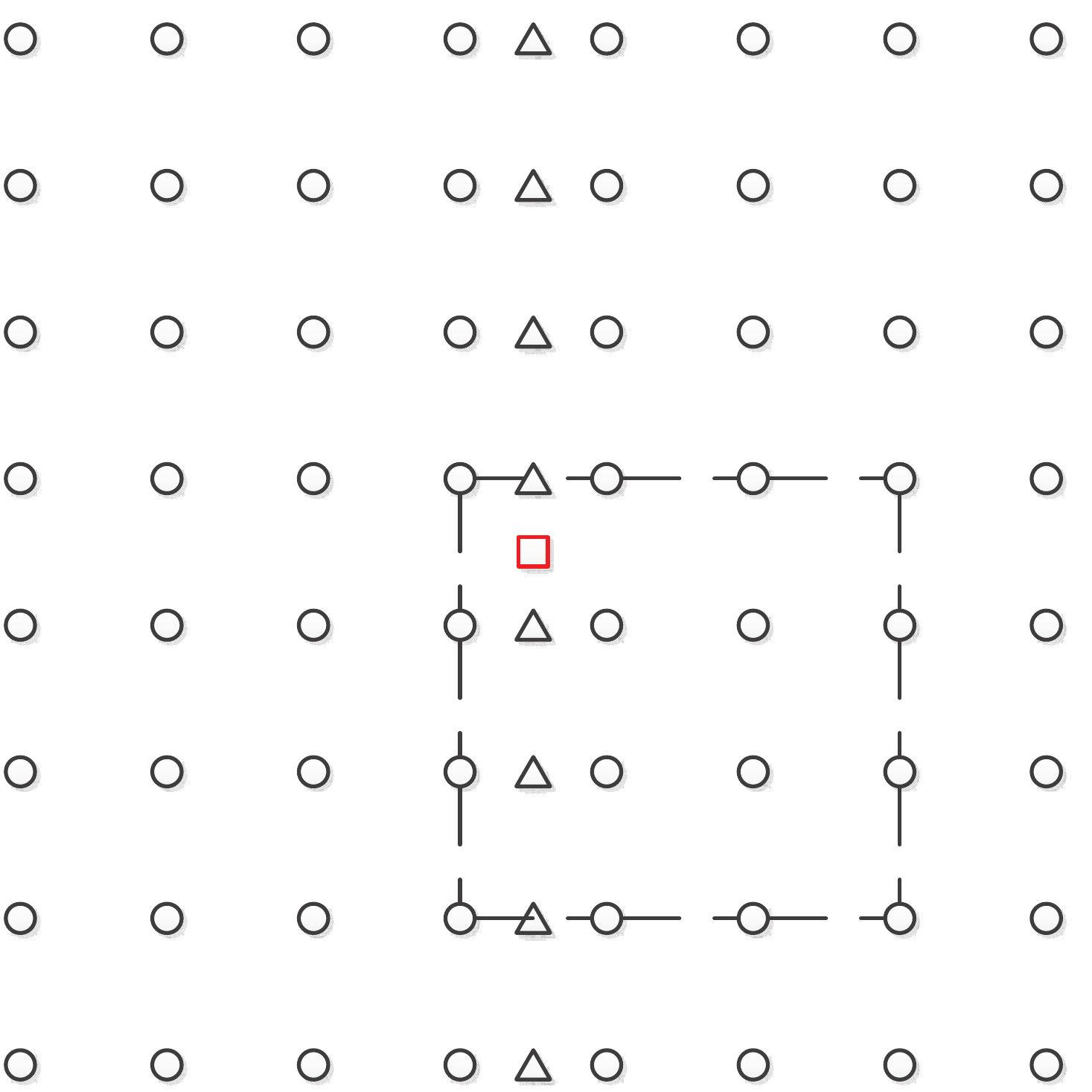}
}
\subfigure[ block-based MC, $44 + 16$ interpolations ]
{
\includegraphics[width=0.4\textwidth]{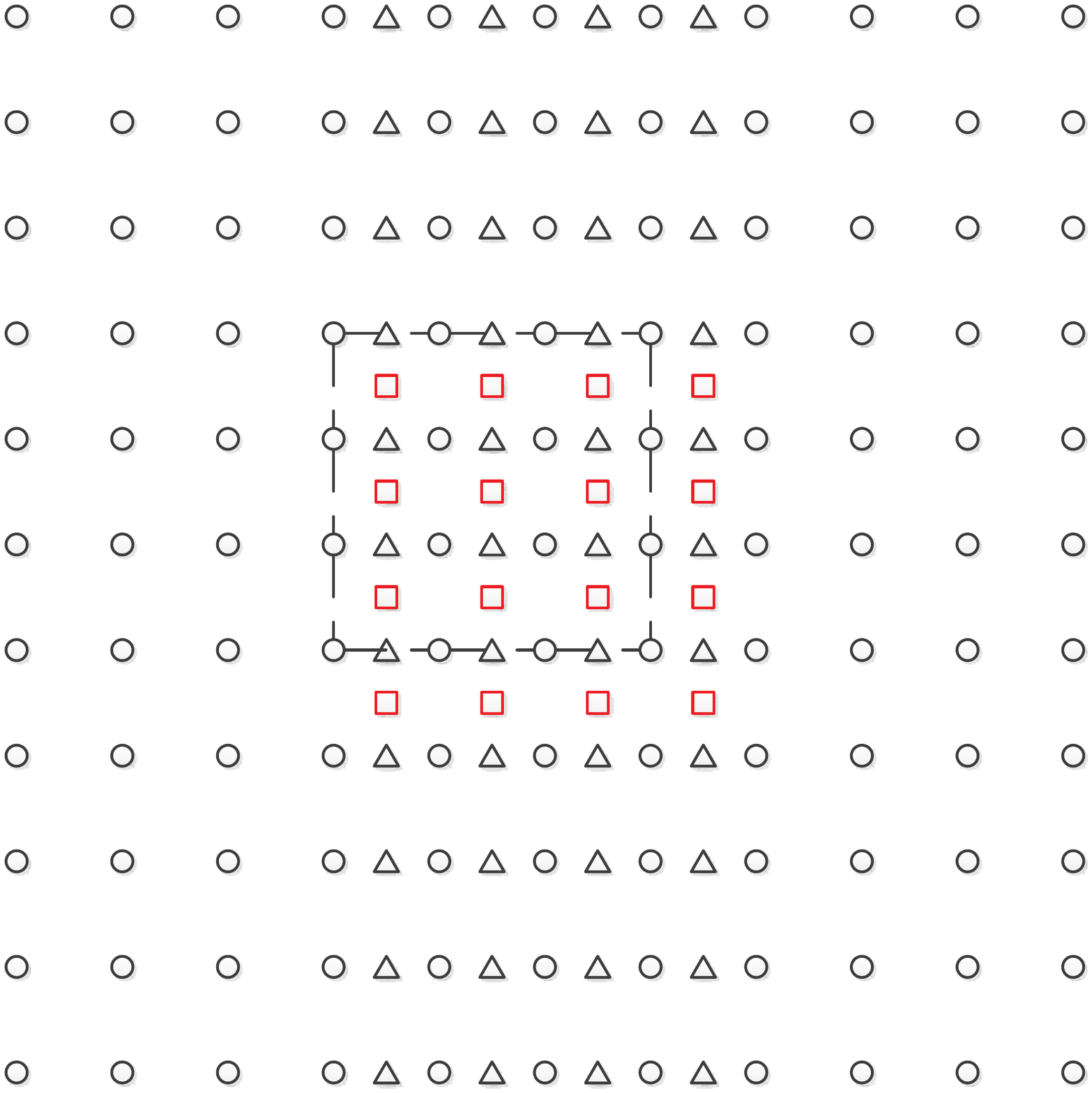}
}
\caption{An example of the influence of MC size}
\label{fig:MC_size}
\end{figure}

To further reduce the MC complexity, we try to decrease the MV resolution from pixel level to block level to decrease the interpolation times.
As shown in Fig. \ref{fig:MC_size} (a), if the size of the MC unit is a pixel, which allows different pixels with various MVs, to interpolate a $(\frac{1}{2}, \frac{1}{2})$ pixel shown as the small red square, $9$ fractional pixels ($8$ horizontal interpolations to obtain the triangles and $1$ vertical to obtain the red square) need to be calculated.
Therefore, to interpolate a $4 \times 4$ block, we need to perform totally $9 \times 16 = 144$ times of interpolation.
However, as shown in Fig. \ref{fig:MC_size} (b), if the size of the MC unit is a $4 \times 4$ block, only totally $60$ fractional pixels ($44$ horizontal interpolations to obtain the triangles, $16$ vertical interpolations to obtain the red squares) need to be calculated.
The reduction of the interpolation times mainly comes from the reduction of repetitive interpolation operations for the neighboring pixels with the same MV.
It should be noted that the difference between block-based MC and pixel-based MC will become much more significant if the MC block size becomes larger.

According to the above analysis, we know that decreasing the MV resolution from the pixel level to the block level can reduce the MC complexity significantly.
However, it may also lead to some R-D performance losses on high-order motion models.
To deal with such a problem, an affine interpolation-precision-based adaptive block size MC scheme is proposed to adapt to the various video characteristics.
The basic concept of the affine interpolation-precision-based adaptive block size MC is that the affine MV precision of the current block will not be lower than a predefined precision.
The size of each MC unit is determined by
\begin{equation}
\label{affine block size}
\begin{split}
MC_{width} &= PU_{width} / MVD_t * prec          \\
MC_{height}& = PU_{height} / MVD_l * prec
\end{split}
\end{equation}
where $PU_{width}$ and $PU_{height}$ are the width and height of the current PU.
All the pixels in a PU share the same affine motion parameters.
$MVD_t$ is the relatively larger MV difference between the horizontal and vertical directions for the top left and top right positions.
$MVD_l$ is the corresponding MV difference of the top left and bottom left positions.
$prec$ determines the minimum precision of the affine interpolation, and it is set as $\frac{1}{8}$ in our experiment.
It can be seen from (\ref{affine block size}) that the larger $MVD_t$ or $MVD_l$ is, the smaller the size of the MC unit will be to guarantee the affine interpolation precision.
The minimum size of an affine MC unit is set as $4$.

\section{Experimental Results}
\label{Sec::experimental results}

\subsection{Simulation setup}
\label{subsec::simulation}
To evaluate the performance of the proposed four-parameter affine MC framework, the proposed algorithm is implemented in the HEVC reference software HM-16.7 \cite{HM167}.
The low delay (LD) main profile and random access (RA) main profile configurations specified in \cite{Frank2013} are used as the test conditions.
The quantization parameters (QP) tested are $22$, $27$, $32$, and $37$ following the HEVC common test condition.
Bjontegaard Delta-rate (BD-rate) \cite{Bjontegaard2001} is employed in our experiments for fair R-D performance comparison.

As the proposed algorithm is designed to better characterize the combination of rotation, zooming, and translation, some sequences with rich rotation or zooming motions are selected to verify the performance of the proposed algorithm.
To be more specific, some segments with rich rotation or zooming motions are extracted from the full sequences to better explain the benefits brought by the proposed algorithm.
These segments will be called as affine test sequences in the following parts.
The detailed characteristics of the affine test sequences are shown in Table \ref{Sequences with affine}.
From Table \ref{Sequences with affine}, we can see that the affine test sequences include various spatial/temporal resolutions and different characteristics.

\begin{table}[tp]
\caption{Characteristics of affine test sequences (sequences with rich rotation or zooming motions)}
\label{Sequences with affine}
\center
\begin{tabular}{rrrrr}
\hline
Sequence        &  Picture        & Video             & Video       & Frame \\
 name           &  order count    & characteristic   & resolution  & rate  \\
\hline
Tractor(TR)       &    591-690     & zooming      & 1920x1080  & 25 \\
Shields(SH)       &    415-514     & zooming      & 1920x1080  & 50 \\
Jets(JE)          &    0-99        & zooming      & 1280x720   & 25 \\
Cactus(CA)        &    0-99        & rotation     & 1920x1080  & 50 \\
BlueSky(BS)       &    0-99        & rotation     & 1920x1080  & 25 \\
Station(ST)       &    0-99        & zooming      & 1920x1080  & 25 \\
SpinCalendar(SC)  &    0-99        & rotation     & 1280x720   & 50 \\
CatRobot(CR)      &    0-99        & rotation     & 3840x2160  & 60 \\
RollerCoaster(RC) &    0-99        & rotation     & 4096x2160  & 60 \\
\hline
\end{tabular}
\end{table}

\subsection{Performance}
\label{subsec::performance}
In this subsection, we will first show the performance of the entire four-parameter affine MC framework for the affine test sequences.
The entire framework contains all the algorithms introduced in this paper including AAMVP combined with the fast ME algorithm, AMM, and the two fast affine MC tools.
Then the performance of the proposed framework will be carefully investigated step by step through the experimental results on affine test sequences in the following three aspects: AAMVP combined with the fast ME algorithm, AMM and AAMVP combined with the fast ME algorithm, and the two fast affine MC coding tools.
Both the R-D performance and encoder/decoder complexity are shown in details.
Finally, we will present some experimental results on the sequences defined in the HEVC common test conditions.

\vspace{3ex}

\subsubsection{The performance of the entire framework on the affine test sequences compared with HM16.7 anchor}
\label{subsubsec::whole framework}

\begin{table}[tp]
\caption{Overall performance of the entire framework compared with HM-16.7 anchor on affine test sequences}
\label{Combine affine result}
\center
\begin{tabular}{r|ccc|ccc}
\hline
Affine         &  \multicolumn{3}{|c}{RA}     & \multicolumn{3}{|c}{LD}        \\
Class             &  Y    & U   & V             & Y    & U   & V                \\
\hline
TR            &    --22.5\%    &    --20.1\%    &    --20.1\%    &    --33.4\%    &    --29.4\%    &    --28.8\%    \\
SH            &    --12.5\%    &    --9.5\%     &    --9.8\%     &    --19.2\%    &    --15.1\%    &    --17.6\%    \\
JE            &    --6.5\%     &    --6.1\%     &    --5.7\%     &    --24.5\%    &    --24.5\%    &    --24.2\%    \\
CA            &    --6.0\%     &    --5.2\%     &    --4.8\%     &    --7.3\%     &    --6.5\%     &    --6.6\%     \\
BS            &    --6.7\%     &    --7.3\%     &    --6.5\%     &    --10.2\%    &    --10.6\%    &    --8.6\%     \\
ST            &    --21.7\%    &    --19.3\%    &    --19.4\%    &    --35.8\%    &    --31.7\%    &    --32.7\%    \\
SC            &    --14.6\%    &    --15.1\%    &    --13.8\%    &    --33.0\%    &    --34.8\%    &    --33.4\%    \\
CR            &    --4.9\%     &    --4.8\%     &    --4.7\%     &    --5.4\%     &    --5.2\%     &    --4.5\%     \\
RC            &    --4.2\%     &    --4.5\%     &    --4.0\%     &    --4.6\%     &    --4.9\%     &    --4.7\%     \\
Avg.           &    --11.1\%    &    --10.2\%    &    --9.9\%     &    --19.3\%    &    --18.1\%    &    --17.9\%    \\
\hline
EncT        &   \multicolumn{3}{|c}{121\%}       &   \multicolumn{3}{|c}{131\%}       \\
DecT        &   \multicolumn{3}{|c}{112\%}       &   \multicolumn{3}{|c}{123\%}       \\
\hline
\end{tabular}
\end{table}

Both the R-D performance and encoding/decoding time increase of the entire affine MC framework for affine test sequences are illustrated in Table \ref{Combine affine result}.
From Table \ref{Combine affine result}, we can see that the proposed framework can achieve averagely $11.1\%$ and $19.3\%$ BD-rate reductions on Y component for the affine test sequences in RA and LD cases, respectively.
The experimental results obviously demonstrate that the proposed four-parameter affine motion model can well represent the combination of rotation, zooming, and translation.

Besides the R-D performance, the encoding/decoding complexity is another important factor to be measured.
From Table \ref{Combine affine result}, we can see that the encoding time is about $121\%$ to $131\%$, and the decoding time is about $112\%$ to $123\%$ compared with the HM-16.7 anchor.
The experimental results obviously show that the proposed affine MC framework will not increase the encoding and decoding complexity significantly.

\vspace{3ex}

\subsubsection{AAMVP combined with the fast Affine ME}
\label{subsubsec::AAMVP}

\begin{table}[tp]
\caption{The overall performance of only enabling the AAMVP combined with the fast affine ME compared with HM-16.7 anchor on affine test sequences}
\label{AAMVP affine result}
\center
\begin{tabular}{r|ccc|ccc}
\hline
Affine         &  \multicolumn{3}{|c}{RA}     & \multicolumn{3}{|c}{LD}        \\
Class             &  Y    & U   & V             & Y    & U   & V                \\
\hline
TR    &    --22.5\%    &    --19.3\%    &    --19.1\%    &    --31.0\%    &    --27.0\%    &    --26.1\%    \\
SH    &    --10.4\%    &    --8.8\%     &    --8.9\%     &    --14.2\%    &    --12.1\%    &    --13.3\%    \\
JE    &    --4.6\%     &    --4.8\%     &    --4.6\%     &    --18.7\%    &    --19.9\%    &    --18.3\%    \\
CA    &    --7.2\%     &    --6.2\%     &    --5.4\%     &    --7.8\%     &    --7.4\%     &    --7.4\%    \\
BS    &    --6.8\%     &    --6.7\%     &    --6.3\%     &    --9.8\%     &    --8.4\%     &    --8.6\%    \\
ST    &    --19.8\%    &    --16.8\%    &    --16.7\%    &    --30.8\%    &    --26.9\%    &    --27.4\%    \\
SC    &    --14.0\%    &    --13.6\%    &    --12.5\%    &    --29.5\%    &    --32.3\%    &    --30.6\%    \\
CR    &    --6.0\%     &    --4.9\%     &    --4.5\%     &    --5.6\%     &    --3.9\%     &    --3.4\%     \\ % to be checked
RC    &    --3.2\%     &    --3.1\%     &    --2.7\%     &    --3.5\%     &    --3.3\%     &    --3.5\%     \\
Avg.  &    --10.5\%    &    --9.4\%     &    --9.0\%     &    --16.8\%    &    --15.7\%    &    --15.4\%    \\
\hline
EncT        &   \multicolumn{3}{|c}{232\%}       &   \multicolumn{3}{|c}{296\%}       \\
DecT        &   \multicolumn{3}{|c}{250\%}       &   \multicolumn{3}{|c}{409\%}       \\
\hline
\end{tabular}
\end{table}

The overall R-D performance and encoding/decoding time increase of only enabling the AAMVP combined with the fast affine ME for affine test sequences compared with HM-16.7 anchor are illustrated in Table \ref{AAMVP affine result}.
From Table \ref{AAMVP affine result}, we can see that the proposed AAMVP combined with the fast affine ME algorithm can provide averagely $10.5\%$ and $16.8\%$ bitrate reductions on Y component in RA and LD cases, respectively.
From the results in Table \ref{AAMVP affine result} and Table \ref{Combine affine result}, we can obviously see that the AAMVP mode contributes most of the BD-rate reductions provided by the four-parameter affine MC framework.
For the encoding/decoding time, from Table \ref{AAMVP affine result}, we can see that the encoding time is about $232\%$ to $296\%$ compared with the HM-16.7 anchor.
Besides, the decoding time is about $250\%$ to $409\%$ accordingly.
Both the encoding and decoding burdens are quite high since the fast affine MC coding tools are not used in the test.

\vspace{3ex}

\subsubsection{AMM and AAMVP combined with the fast ME algorithm}
\label{subsubsec::AMM}
The AMM mode always attempts to reuse the affine motion model of the neighboring blocks using affine mode.
Therefore, the AMM mode cannot be used without AAMVP.
In this subsection, we will see both the R-D performance improvement and encoding/decoding time change brought by the AMM mode on the premise of AAMVP.

The average bitrate savings and encoding/decoding time increase of the AMM and AAMVP combined with the fast affine ME algorithm for affine test sequences compared with HM-16.7 anchor are shown in Table \ref{AMM affine result}.
Through the comparison between Table \ref{AMM affine result} and Table \ref{AAMVP affine result}, we can see that the proposed AMM mode can further achieve about $1.7\%$ and $3.0\%$ R-D performance improvements in average on the premise of AAMVP mode in RA and LD cases, respectively.
The experimental results demonstrate that the proposed AMM mode is beneficial to the overall R-D performance.
As for the encoding/decoding complexity, the encoding time of the AMM combined with AAMVP only increases a little compared with the AAMVP mode since the AMM mode without complex affine ME operations is quite simple.
However, the decoding time increases very obviously.
The reason is that the number of blocks choosing affine mode increases significantly, and then the computations of affine MC are multiplied.
The average decoding time for the affine test sequences can be as much as $523\%$ compared with the HEVC anchor in LD case.
Note that in this test, the fast affine MC coding tools are still not used.

\begin{table}[tp]
\caption{Overall performance of enabling the AMM and AAMVP combined with the fast affine ME compared with HM-16.7 anchor on affine test sequences}
\label{AMM affine result}
\center
\begin{tabular}{r|ccc|ccc}
\hline
Affine         &  \multicolumn{3}{|c}{RA}     & \multicolumn{3}{|c}{LD}        \\
Class             &  Y    & U   & V             & Y    & U   & V                \\
\hline
TR    &    --24.0\%    &    --21.0\%    &    --21.0\%    &    --34.7\%    &    --31.4\%    &    --30.2\%    \\
SH    &    --13.5\%    &    --11.1\%    &    -11.1\%     &    --19.7\%    &    --17.7\%    &    --18.9\%    \\
JE    &    --6.1\%     &    --5.4\%     &    --5.2\%     &    --23.5\%    &    --24.4\%    &    --23.2\%    \\
CA    &    --8.0\%     &    --6.7\%     &    --6.1\%     &    --8.7\%     &    --8.3\%     &    --8.3\%    \\
BS    &    --8.1\%     &    --7.8\%     &    --7.2\%     &    --11.1\%    &    --10.1\%    &    --9.8\%    \\
ST    &    --22.7\%    &    --19.6\%    &    --19.7\%    &    --37.1\%    &    --32.8\%    &    --33.6\%    \\
SC    &    --15.8\%    &    --16.1\%    &    --14.4\%    &    --32.5\%    &    --35.5\%    &    --33.9\%    \\
CR    &    --6.7\%     &    --5.9\%     &    --5.3\%     &    --6.0\%     &    --4.8\%     &    --4.1\%     \\
RC    &    --4.7\%     &    --4.7\%     &    --4.3\%     &    --4.7\%     &    --4.9\%     &    --4.6\%     \\
Avg.  &    --12.2\%    &    --10.9\%    &    --10.5\%    &    --19.8\%    &    --18.9\%    &    --18.5\%    \\
\hline
EncT        &   \multicolumn{3}{|c}{236\%}       &   \multicolumn{3}{|c}{301\%}       \\
DecT        &   \multicolumn{3}{|c}{326\%}       &   \multicolumn{3}{|c}{523\%}       \\
\hline
\end{tabular}
\end{table}

\vspace{3ex}

\subsubsection{The performance of one-step interpolation filter}
\label{subsubsec::one-step}
In this subsubsection, the performances of the proposed two fast affine MC coding tools are presented.
The overall R-D performance and encoding/decoding time change of the one-step sub-pixel interpolation filter for affine test sequences compared with HM-16.7 anchor are shown in Table \ref{Unifilter affine result}.
Through the comparison between Table \ref{AMM affine result} and Table \ref{Unifilter affine result}, we can see that the DCTIF based one-step sub-pixel interpolation filter can further bring about $0.3\%$ and $0.9\%$ bitrate savings averagely on Y component for affine test sequences in RA and LD cases compared with the situation using bilinear interpolation filter.
The benefits mainly come from the fact that the DCTIF can bring better interpolation results than the bilinear interpolation filter when the MV precision is higher than $\frac{1}{4}$ pixel.
Besides, since the one-step sub-pixel interpolation filter can reduce the interpolation times dramatically for those blocks with affine MVs higher than $\frac{1}{4}$ pixel precision, the encoding and decoding complexities are both reduced significantly.
Especially, for the affine test sequences in LD case, the decoding time reduces from $523\%$ to $170\%$.

\begin{table}[tp]
\caption{Overall performance of enabling one-step interpolation filter, AMM, and AAMVP combined with the fast affine ME algorithm compared with HM-16.7 anchor on affine test sequences}
\label{Unifilter affine result}
\center
\begin{tabular}{r|ccc|ccc}
\hline
Affine         &  \multicolumn{3}{|c}{RA}     & \multicolumn{3}{|c}{LD}        \\
Class             &  Y    & U   & V             & Y    & U   & V                \\
\hline
TR    &    --24.0\%    &    --21.1\%    &    --21.0\%    &    --35.0\%    &    --31.4\%    &    --30.3\%    \\
SH    &    --13.6\%    &    --10.7\%    &    --10.5\%    &    --20.1\%    &    --16.9\%    &    --19.0\%    \\
JE    &    --6.8\%     &    --6.2\%     &    --5.6\%     &    --25.4\%    &    --25.8\%    &    --24.2\%    \\
CA    &    --8.4\%     &    --7.1\%     &    --6.7\%     &    --9.2\%     &    --8.4\%     &    --8.4\%    \\
BS    &    --8.3\%     &    --8.7\%     &    --7.7\%     &    --12.0\%    &    --11.6\%    &    --10.1\%    \\
ST    &    --23.0\%    &    --19.6\%    &    --20.0\%    &    --37.3\%    &    --32.1\%    &    --33.9\%    \\
SC    &    --17.0\%    &    --16.6\%    &    --15.7\%    &    --36.3\%    &   --37.5\%     &    --37.1\%    \\
CR    &    --6.8\%     &    --6.0\%     &    --5.7\%     &    --6.2\%     &    --5.4\%     &    --4.6\%     \\
RC    &    --4.7\%     &    --4.6\%     &    --4.2\%     &    --4.8\%     &    --5.0\%     &    --4.9\%     \\
Avg.  &    --12.5\%    &    --11.2\%    &    --10.8\%    &    --20.7\%    &    --19.4\%    &    --19.2\%    \\
\hline
EncT        &   \multicolumn{3}{|c}{136\%}       &   \multicolumn{3}{|c}{153\%}       \\
DecT        &   \multicolumn{3}{|c}{135\%}       &   \multicolumn{3}{|c}{170\%}       \\
\hline
\end{tabular}
\end{table}

\vspace{3ex}

\subsubsection{The performance of adaptive block size MC}
\label{subsubsec::block-MC}
The average performance and encoding/decoding time change of the affine interpolation-precision-based adaptive block size MC for affine test sequences compared with HM-16.7 anchor are shown in Table \ref{ABSM affine result}.
Through the comparison between Table \ref{AMM affine result} and Table \ref{ABSM affine result}, we can see that the BD-rate increases by about $1.4\%$ and $1.5\%$ in RA and LD cases in average for affine test sequences compared with the situation without any fast MC algorithms.
The losses mainly come from the fact that the minimum size of the MC unit is set as $4$ to reduce the computational burden for both the hardware and software implementation.
Although the adaptive block size MC will incur a few performance losses, it can bring quite a significant encoding/decoding complexity reduction.
As an example, the decoding time decreases from $523\%$ to $199\%$ for affine test sequences in LD case.

\begin{table}[tp]
\caption{Overall performance of enabling the adaptive block size MC, AMM, and AAMVP combined with the fast affine ME algorithm compared with HM-16.7 anchor on affine test sequences}
\label{ABSM affine result}
\center
\begin{tabular}{r|ccc|ccc}
\hline
Affine         &  \multicolumn{3}{|c}{RA}     & \multicolumn{3}{|c}{LD}        \\
Class             &  Y    & U   & V             & Y    & U   & V                \\
\hline
TR    &    --22.4\%    &    --20.1\%    &    --19.8\%    &    --33.0\%    &    --28.5\%    &    --28.3\%    \\
SH    &    --12.5\%    &    --10.3\%    &    --10.3\%    &    --18.9\%    &    --16.5\%    &    --18.6\%    \\
JE    &    --6.0\%     &    --5.3\%     &    --5.3\%     &    --22.4\%    &    --23.7\%    &    --22.5\%    \\
CA    &    --5.7\%     &    --5.0\%     &    --4.5\%     &    --6.7\%     &    --6.6\%     &    --6.2\%    \\
BS    &    --6.5\%     &    --6.4\%     &    --5.9\%     &    --9.1\%     &    --8.5\%     &    --7.5\%    \\
ST    &    --21.4\%    &    --19.2\%    &    --19.0\%    &    --35.6\%    &    --32.2\%    &    --31.9\%    \\
SC    &    --13.2\%    &    --13.9\%    &    --12.6\%    &    --29.2\%    &    --32.3\%    &    --30.9\%    \\
CR    &    --5.0\%     &    --4.8\%     &    --4.5\%     &    --5.3\%     &    --4.7\%     &    --4.2\%     \\
RC    &    --4.3\%     &    --4.7\%     &    --4.0\%     &    --4.6\%     &    --4.5\%     &    --4.3\%     \\
Avg.  &    --10.8\%    &    --10.0\%    &    --9.5\%     &    --18.3\%    &    --17.5\%    &    --17.1\%    \\
\hline
EncT        &   \multicolumn{3}{|c}{138\%}       &   \multicolumn{3}{|c}{156\%}       \\
DecT        &   \multicolumn{3}{|c}{154\%}       &   \multicolumn{3}{|c}{199\%}       \\
\hline
\end{tabular}
\end{table}

\vspace{3ex}

\subsubsection{The performance of two fast affine MC coding tools together}
\label{subsubsec::combine-MC}
The overall R-D performance and encoding/decoding complexity of the combined algorithm for the affine test sequences compared with HM-16.7 anchor have already been shown in Table \ref{Combine affine result} in advance.
Compared with the situation without any fast MC algorithms as shown in Table \ref{AMM affine result}, the fast affine MC algorithms totally just suffer about $1.1\%$ and $0.5\%$ R-D performance losses for the affine test sequences in RA and LD cases, accordingly.
Although there are a few performance losses, the encoding/decoding complexity decrease is quite amazing.
The encoding time is only about $20\%$ to $30\%$ increase compared with the HM-16.7 anchor without affine mode.
The decoding time increases about $12\%$ to $26\%$ for affine test sequences.
The experimental results obviously demonstrate that the combination of the two fast affine MC coding tools can lead to a better trade-off between the R-D performance and encoding/decoding complexity.

\vspace{3ex}

\subsubsection{R-D curve}
\label{RD curve}
\begin{figure*}[tp]
\centering
\subfigure
{
\includegraphics[width=0.48\textwidth]{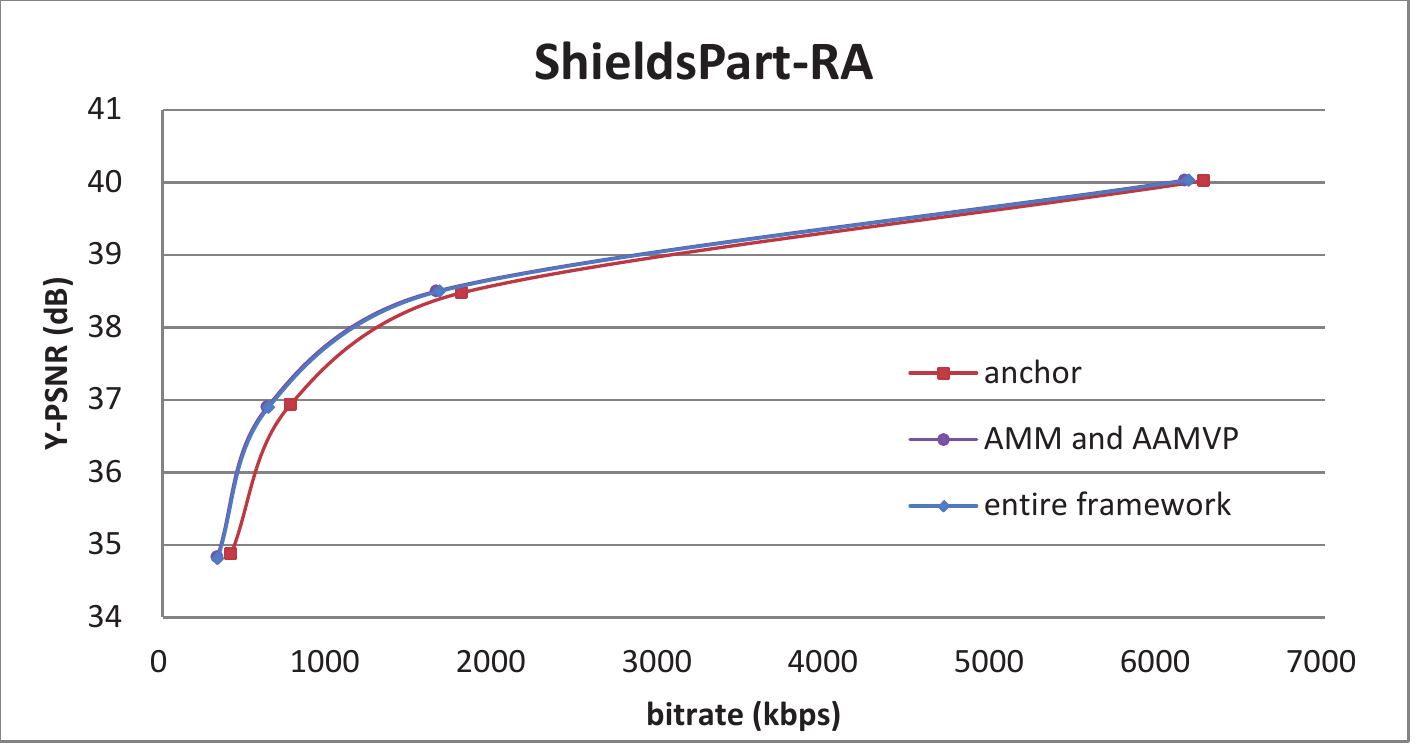}
}
\subfigure
{
\includegraphics[width=0.48\textwidth]{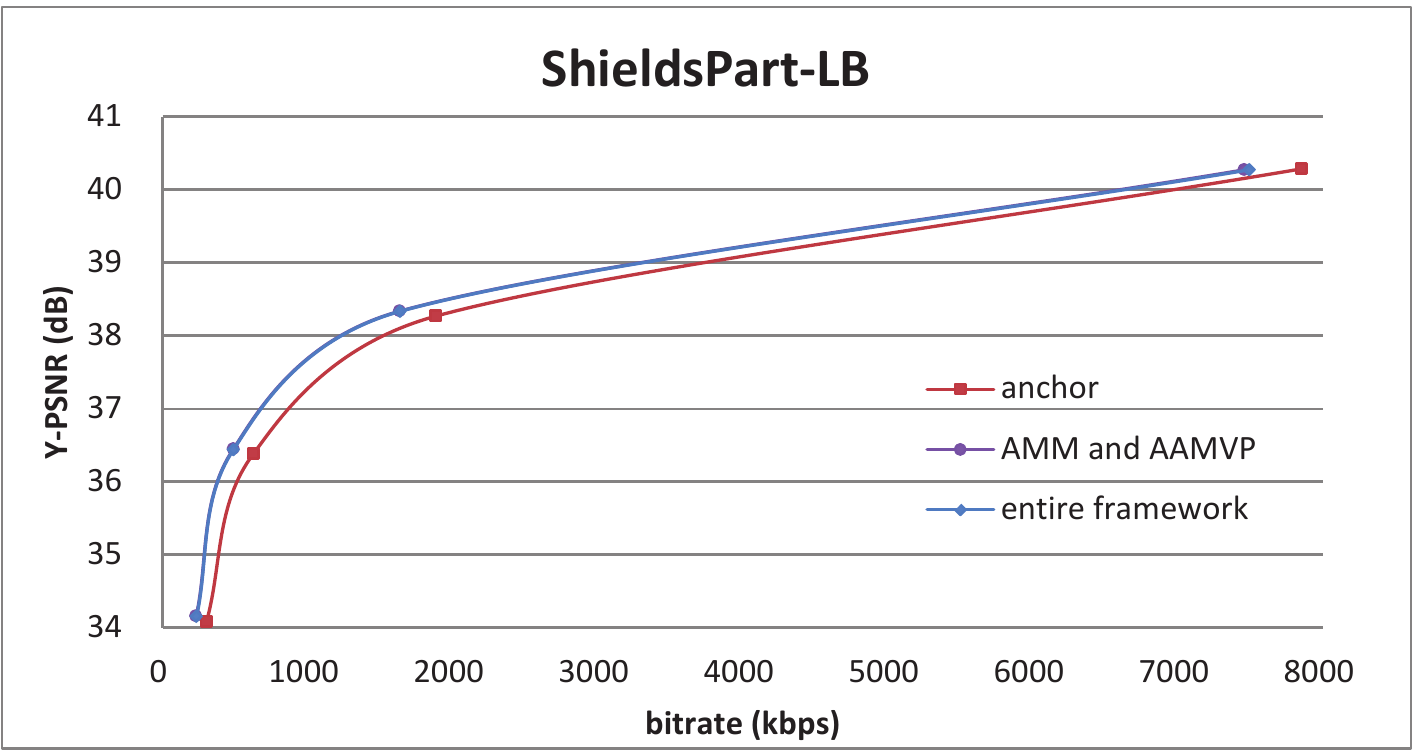}
}
\subfigure
{
\includegraphics[width=0.48\textwidth]{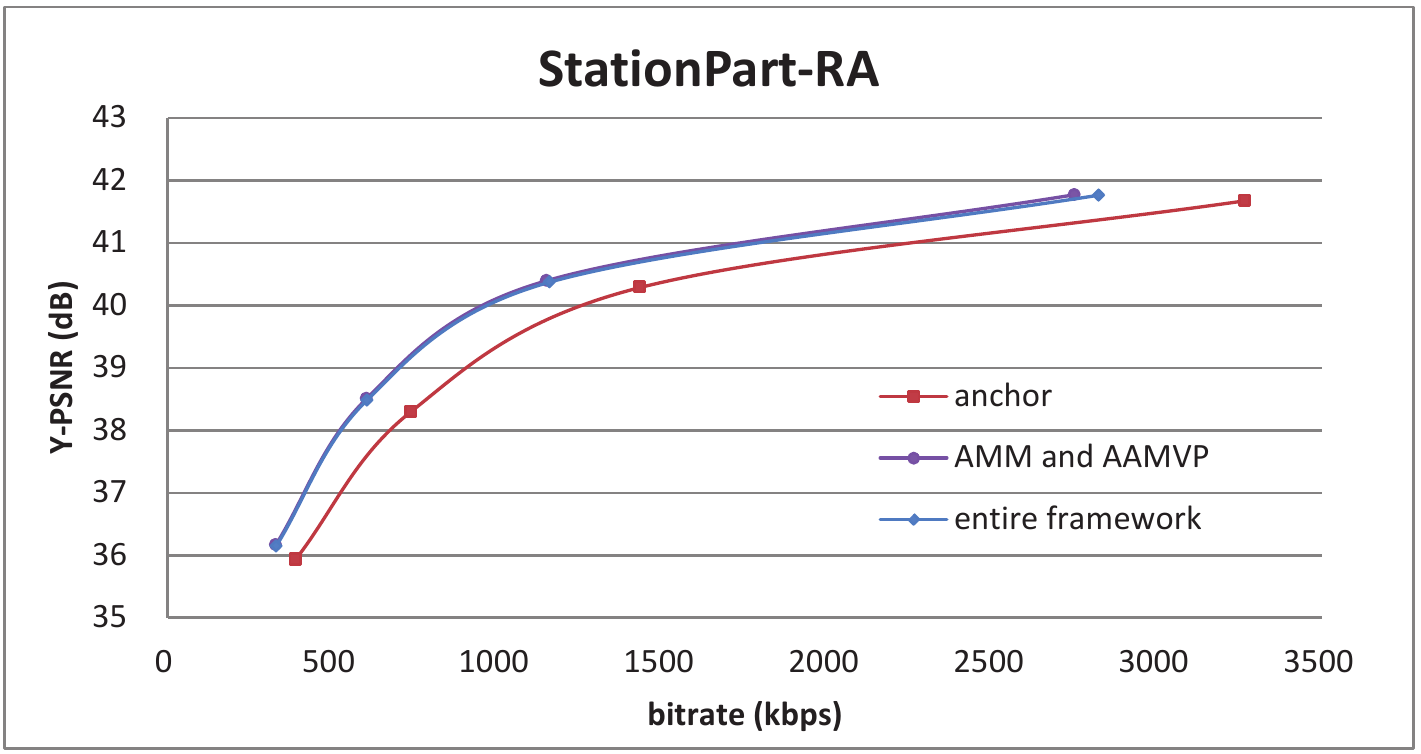}
}
\subfigure
{
\includegraphics[width=0.48\textwidth]{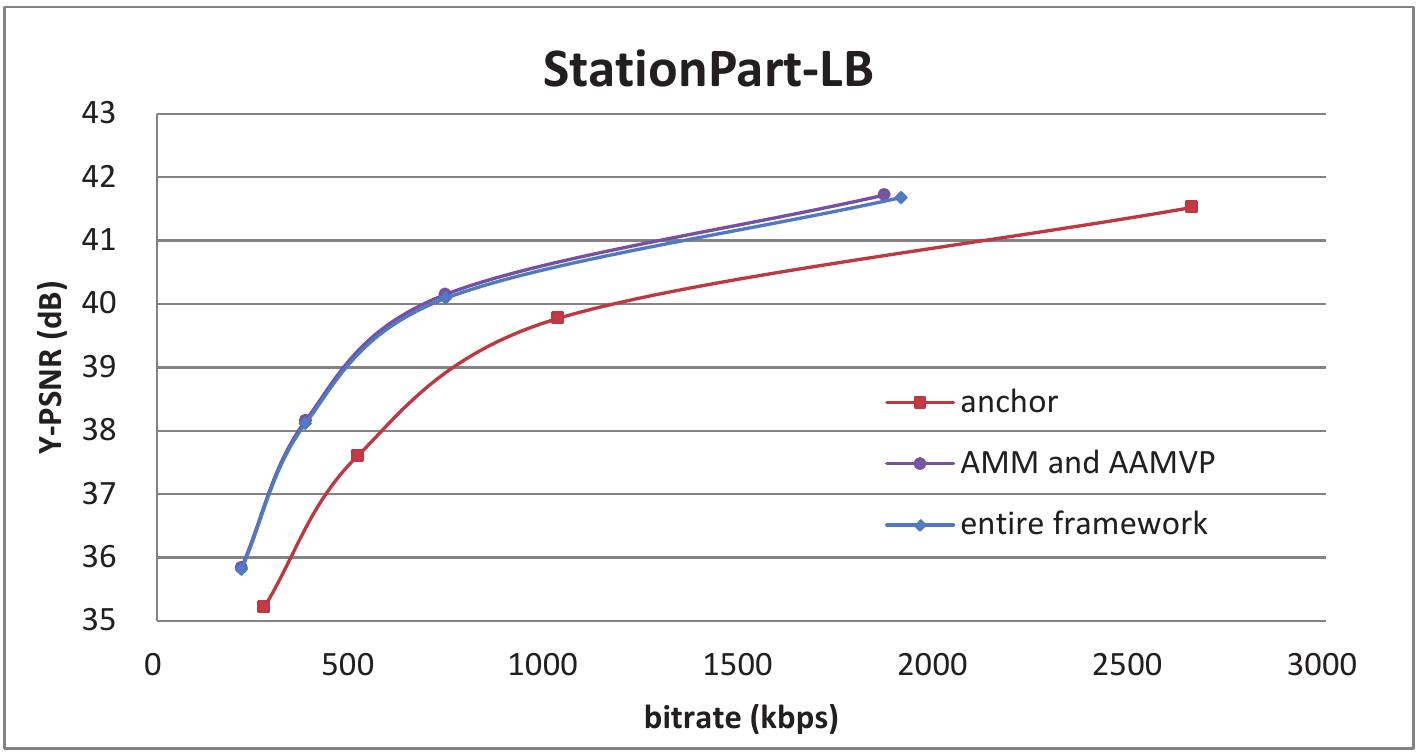}
}
\caption{Some example R-D curves of the proposed affine MC framework}
\label{R-D curve example}
\end{figure*}

Some example R-D curves are shown in Fig. \ref{R-D curve example}.
These R-D curves also obviously demonstrate that the proposed algorithm can achieve much better R-D performance improvement compared with the HEVC anchor.
It can be obviously seen from Fig. \ref{R-D curve example} that the performance improvement mainly comes from the bitrate reduction instead of Y-PSNR improvement.
Besides, we can also see that the entire framework presents ignorable R-D performance losses compared with the ``AMM and AAMVP" (without fast MC coding tools) in the figure.

\vspace{3ex}

\subsubsection{The performance of the entire framework on the affine test sequences compared with the previous work}
\label{subsubsec::whole framework previous}

\begin{table}[tp]
\caption{Overall performance of the previous work \cite{Li2015} compared with HEVC anchor on affine test sequences}
\label{previous affine result}
\center
\begin{tabular}{r|ccc|ccc}
\hline
Affine         &  \multicolumn{3}{|c}{RA}     & \multicolumn{3}{|c}{LD}        \\
Class             &  Y    & U   & V             & Y    & U   & V                \\
\hline
TR            &    --14.6\%    &    --13.4\%    &    --13.1\%    &    --22.5\%    &    --20.1\%    &    --20.1\%    \\
SH            &    --3.4\%     &    --2.9\%     &    --3.2\%     &    --12.5\%    &    --9.5\%     &    --9.8\%     \\
JE            &    --2.3\%     &    --2.0\%     &    --1.9\%     &    --6.5\%     &    --6.1\%     &    --5.7\%     \\
CA            &    --3.5\%     &    --2.9\%     &    --2.8\%     &    --6.0\%     &    --5.2\%     &    --4.8\%     \\
BS            &    --3.1\%     &    --3.6\%     &    --3.1\%     &    --6.7\%     &    --7.3\%     &    --6.5\%     \\
ST            &    --11.2\%    &    --11.0\%    &    --10.7\%    &    --21.7\%    &    --19.3\%    &    --19.4\%    \\
SC            &    --7.3\%     &    --7.3\%     &    --6.2\%     &    --14.6\%    &    --15.1\%    &    --13.8\%     \\
CR            &    --2.0\%     &    --1.4\%     &    --1.8\%     &    --3.6\%     &    --1.8\%     &    --2.0\%     \\
RC            &    --1.8\%     &    --1.6\%     &    --1.5\%     &    --1.0\%     &    --1.6\%     &    --1.2\%     \\
Avg.          &    --5.5\%     &    --5.1\%     &    --5.0\%     &    --6.8\%     &    --5.1\%     &    --4.7\%     \\
\hline
EncT        &   \multicolumn{3}{|c}{419\%}       &   \multicolumn{3}{|c}{323\%}       \\
DecT        &   \multicolumn{3}{|c}{507\%}       &   \multicolumn{3}{|c}{951\%}       \\
\hline
\end{tabular}
\end{table}

We also compare the entire framework with our previous work \cite{Li2015} to demonstrate the effectiveness of the proposed algorithm in this paper.
Table \ref{previous affine result} gives the detailed experimental results of our previous work on the affine test sequences.
From Table \ref{previous affine result}, we can see that the proposed algorithm can bring averagely $5.5\%$ and $6.8\%$ R-D performance improvements compared with the HEVC anchor for the affine test sequences.
Compared with the results of the entire framework shown in Table \ref{Combine affine result}, we can see that the proposed algorithm can bring over double BDrate-savings compared with the previous work.
The significant bitrate savings of the proposed method mainly comes from the following five aspects.
\begin{itemize}
\item The reduction of the parameters from 6 to 4 can effectively reduce the header information so as to improve the R-D performance.
\item In the proposed work in this paper, we have searched over all the reference frames in list0 and list1, and the combination of them to obtain better prediction block. However, in the previous work, we have not focused on developing any fast affine MC algorithms. Therefore, to reduce encoding complexity, we only search the inter direction and the reference frame, which are the same as those of the top left block.
\item The AAMVP can achieve better MVP tuple to accurately predict the affine MV.
\item The AMM mode can save lots of header information.
\item The DCTIF based one step interpolation filter can lead to better coding efficiency compared with the bilinear interpolation filter used in the previous work.
\end{itemize}
Besides, from the encoding/decoding complexity point of view, the proposed algorithm also achieves much lower encoding and decoding complexity compared with the previous work.

\vspace{3ex}

\subsubsection{Performance of the HEVC common test condition sequences}
\label{subsebsec:general test}

\begin{table}[tp]
\caption{Overall performance of the combined algorithm compared with HM-16.7 anchor on HEVC common test condition sequences}
\label{Combine common result}
\center
\begin{tabular}{r|ccc|ccc}
\hline
General          &  \multicolumn{3}{|c}{RA}     & \multicolumn{3}{|c}{LD}        \\
Class             &  Y    & U   & V             & Y    & U   & V                \\
\hline
Class A          &  --0.5\%  &  --0.4\%  &  --0.4\%  &   --      &   --      &   --      \\
Class B          &  --1.4\%  &  --1.2\%  &  --1.1\%  &  --1.5\%  &  --1.5\%  &  --1.3\%  \\
Class C          &  --0.7\%  &  --0.7\%  &  --0.9\%  &  --1.0\%  &  --1.0\%  &  --1.3\%  \\
Class D          &  --1.1\%  &  --1.2\%  &  --1.3\%  &  --2.0\%  &  --2.2\%  &  --2.8\%  \\
Class E          &     --    &   --      &   --      &  --1.8\%  &  --1.6\%  &  --2.0\%  \\
Class F          &  --1.3\%  &  --1.5\%  &  --1.5\%  &  --1.5\%  &  --1.6\%  &  --1.9\%  \\
Avg.             &  --1.0\%  &  --0.9\%  &  --0.9\%  &  --1.5\%  &  --1.6\%  &  --1.8\%  \\
\hline
EncT        &   \multicolumn{3}{|c}{118\%}       &   \multicolumn{3}{|c}{128\%}       \\
DecT        &   \multicolumn{3}{|c}{103\%}       &   \multicolumn{3}{|c}{105\%}       \\
\hline
\end{tabular}
\end{table}

We also present some experimental results on the HEVC common test condition sequences as shown in Table \ref{Combine common result}.
From Table \ref{Combine common result}, we can obviously see that the proposed framework can bring averagely $1.0\%$ and $1.5\%$ R-D performance improvements in RA and LD cases compared with the HM-16.7 anchor for the HEVC common test condition sequences.
Besides, the increase of the complexity of the HEVC common test condition sequences, especially the decoding time, is only marginal.
Therefore, the proposed technique is promising to be integrated into future video coding standards.

\subsection{Performance analysis}
\label{subsec::performance analysis}
In this subsection, the benefits brought by the proposed affine MC framework will be analyzed carefully from the change of the following two factors before and after the use of the affine motion model: the coding unit (CU) partition size, the number of blocks using affine mode.

\begin{figure}[tp]
\centering
{
\includegraphics[width=0.48\textwidth]{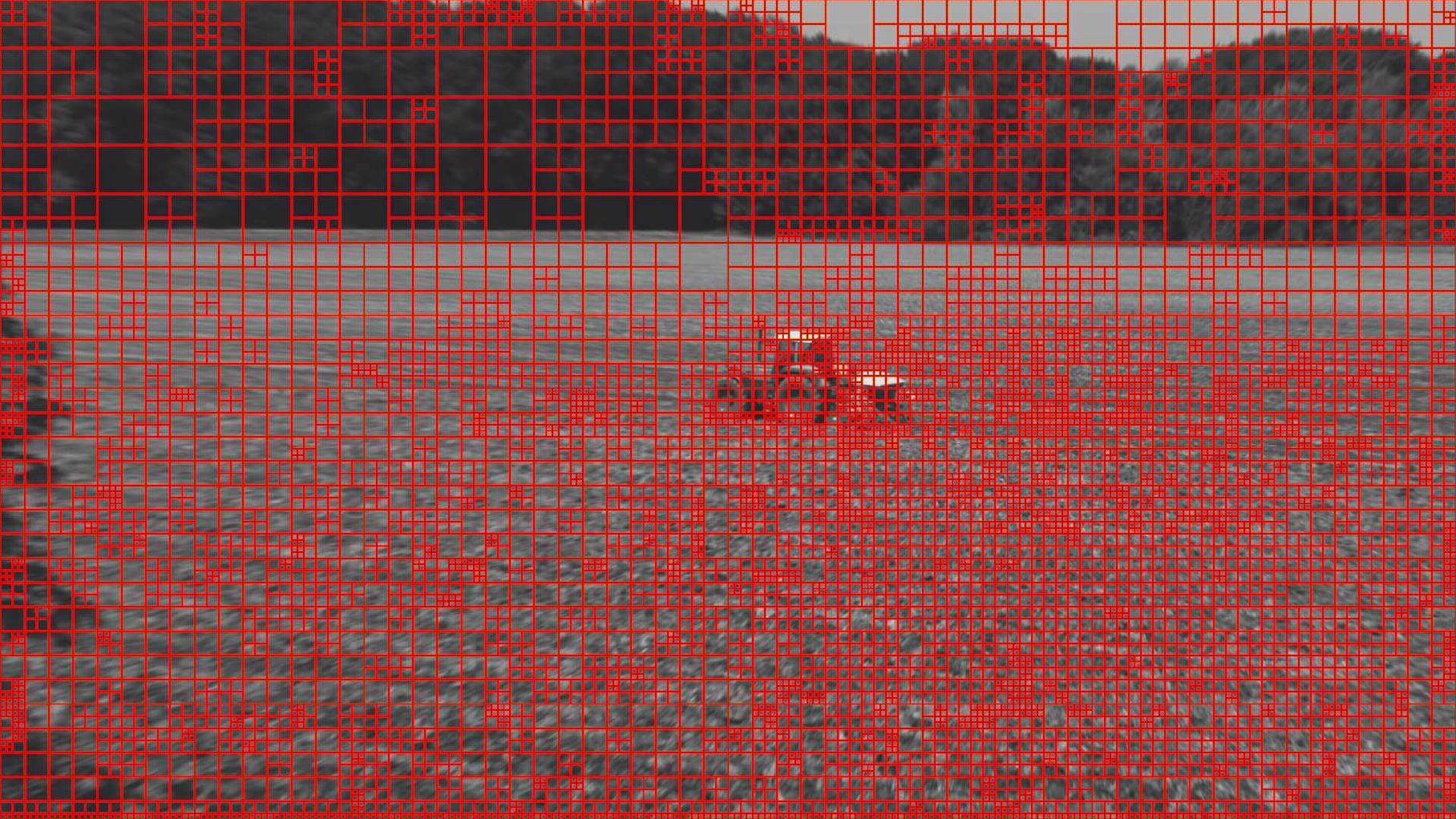}
}
\caption{CU partition of HEVC anchor, tractor, LD, QP27, POC12}
\label{anchor_partition}
\end{figure}

\begin{figure}[tp]
\centering
{
\includegraphics[width=0.48\textwidth]{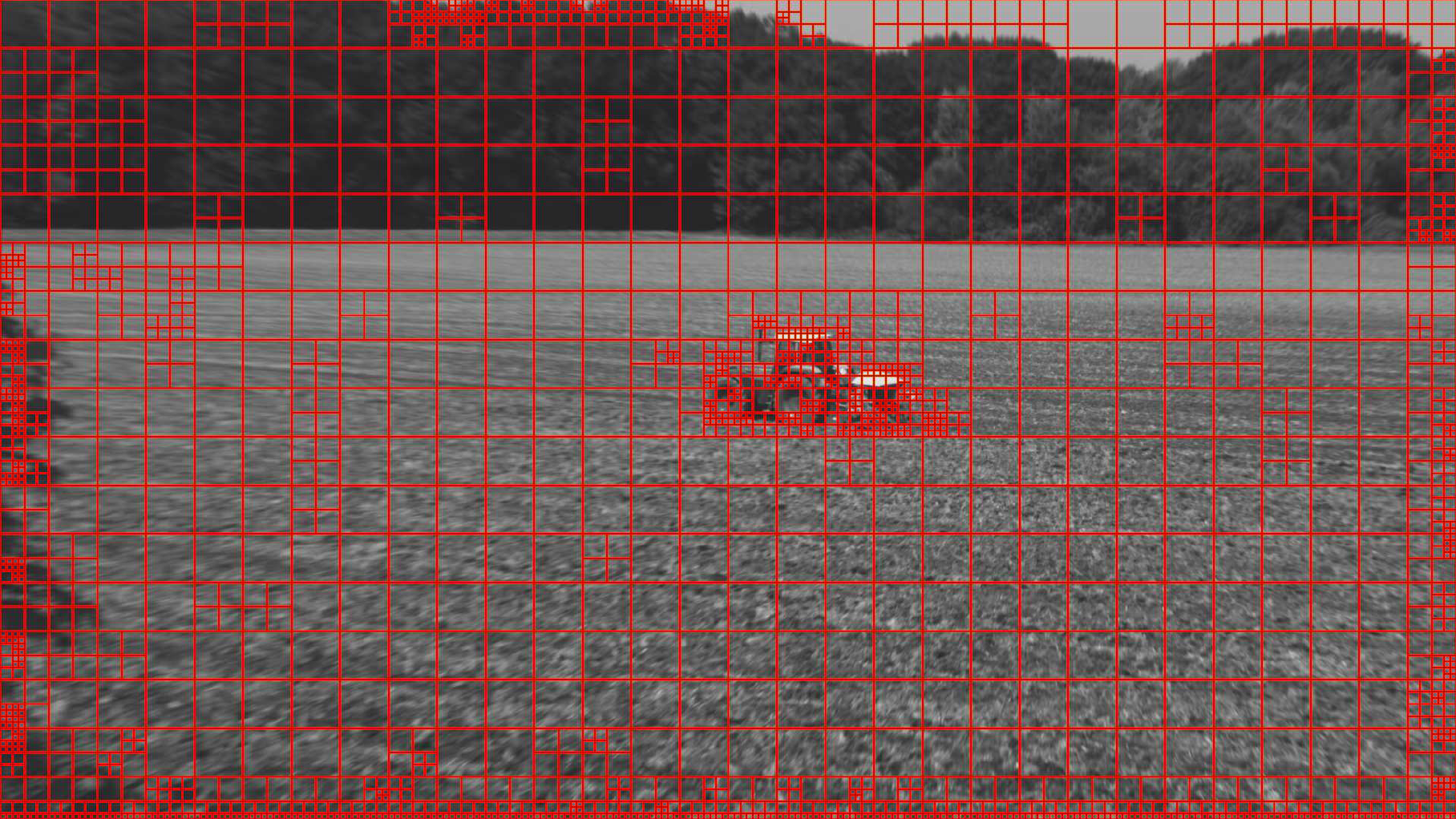}
}
\caption{CU partition of affine mode, tractor, LD, QP27, POC12}
\label{affine partition}
\end{figure}

Fig. \ref{anchor_partition} and Fig. \ref{affine partition} show the CU partition of an inter frame of a typical affine sequence for HEVC anchor and the proposed four-parameter affine motion model framework, respectively.
In both figures, a red square represents a CU and picture order count (POC) means the frame number in display order.
From these two figures, we can obviously see that the CU partitions are quite small for most of the blocks for HEVC anchor while the CU partitions become quite large when the affine motion model is applied.
The reason is that the zooming motion in the sequence can be well represented by the proposed affine motion model and thus large CU partitions can be used.
However, for the HEVC anchor with only translational motion model, the encoder has to split a block into smaller ones so that for each smaller block the motion is approximately translational.
Therefore, using our proposed affine motion model can enable the use of large blocks and thus reduce the overhead bits on block partitions significantly.

\begin{figure}[tp]
\centering
{
\includegraphics[width=0.48\textwidth]{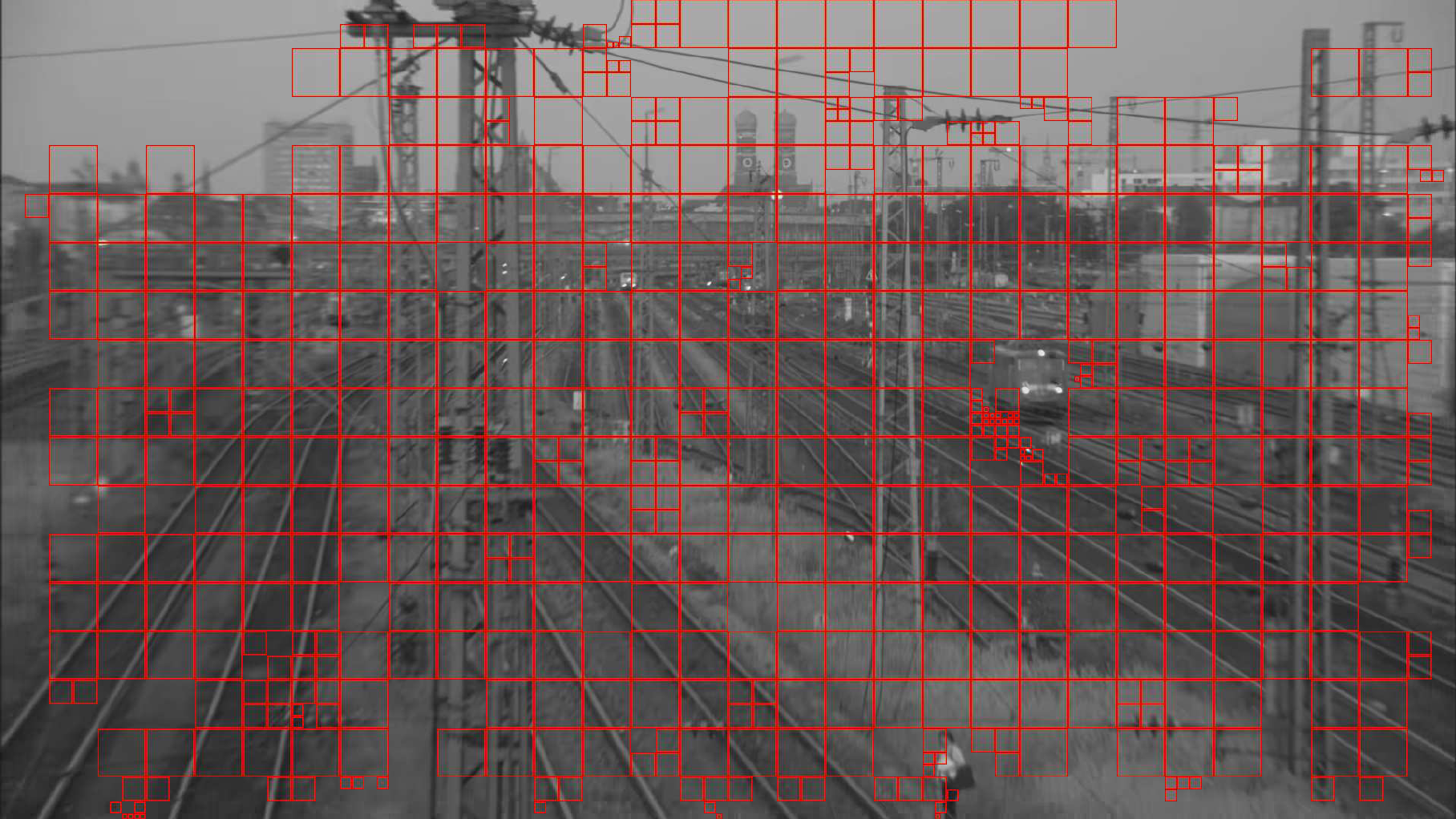}
}
\caption{global affine motion, StationPart, LD, QP32, POC8}
\label{global motion}
\end{figure}

\begin{figure}[tp]
\centering
{
\includegraphics[width=0.48\textwidth]{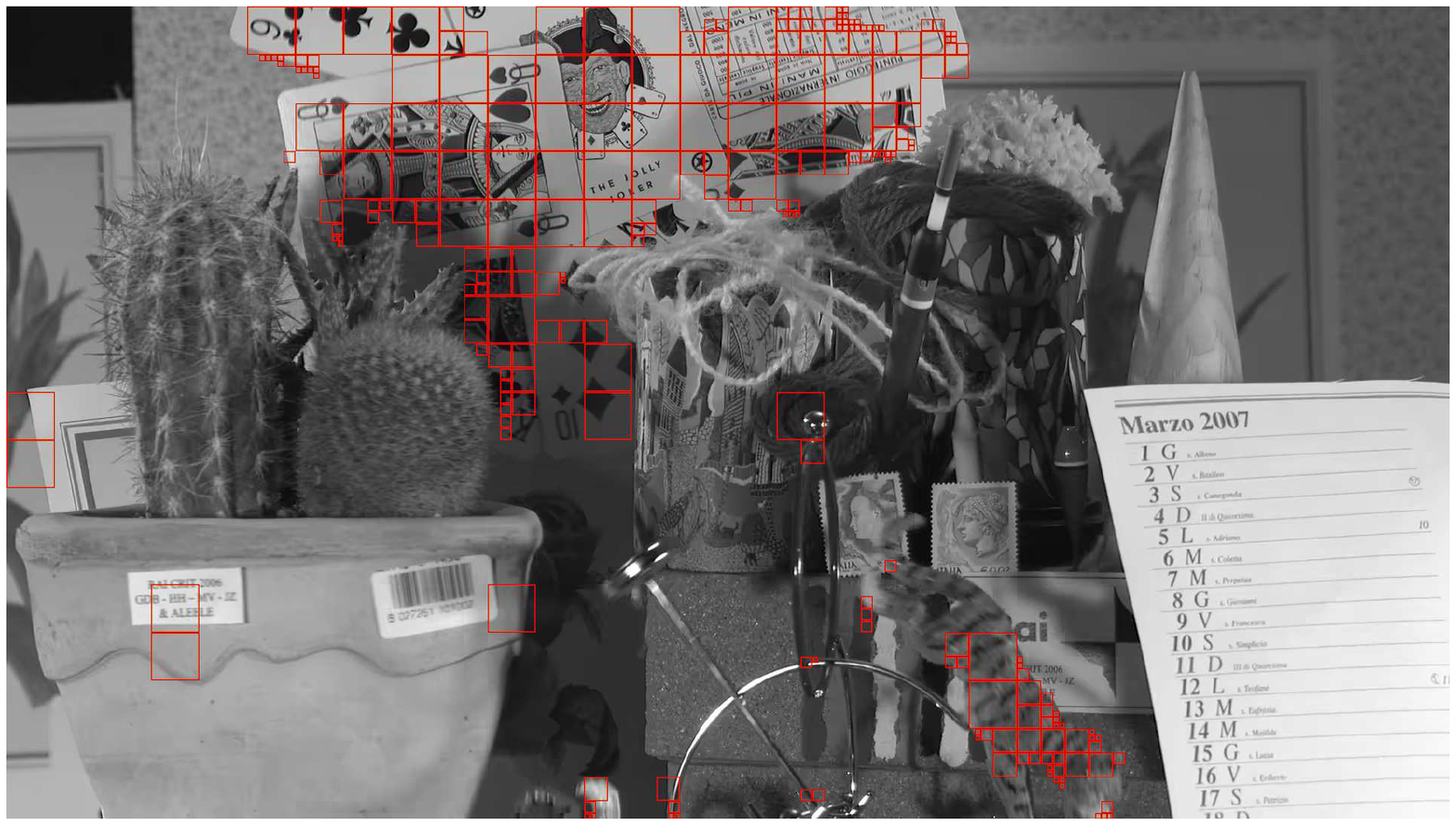}
}
\caption{local affine motion, CactusPart, LD, QP32, POC1}
\label{local motion}
\end{figure}

Fig. \ref{global motion} and Fig. \ref{local motion} show the situations of the blocks using affine motion model in the sequences with global and local affine motion, respectively.
In both figures, the red squares represent the CUs using affine mode.
In the sequence with global affine motion, almost all the blocks choose the affine mode, with only two kinds of blocks being exceptions.
The first kind is the blocks in the border of a frame for which a good prediction is impossible to be obtained.
The second kind is the very smooth blocks such as blue sky for which using translational motion model can also lead to a good prediction.
For the sequence with local affine motion, most of the blocks with rotation motion choose the affine mode while the other blocks with translational motion choose the translation motion model.
The encoder can determine a suitable motion model for each block through RDO.
The experimental results obviously demonstrate that the proposed affine MC framework combined with the traditional translational MC framework can well represent various video contents with global or local complex motions.

Table \ref{affine mode percent} gives the average affine mode percentages for all the frames for the affine test sequences.
It can be seen from Table \ref{affine mode percent} that the affine mode percentages can be quite high for the affine test sequences.
Also, we can see from Table \ref{affine mode percent} that the affine mode percentages in RA case are obviously lower than those in LD case.
This is mainly due to the fact that the utilization of both the forward and backward reference frames is beneficial for obtaining a better prediction for the blocks with complex motions.
Therefore, the benefits brought by the proposed affine motion model in RA case are less than those in LD case.

\begin{table}[tp]
\caption{Affine mode percentage for the affine test sequences with rich rotation or zooming motions}
\label{affine mode percent}
\center
\begin{tabular}{r|c|c}
\hline
Affine Class    &  RA     & LD        \\
\hline
Tractor       &    30.0\%    &    53.7\%    \\
Shields       &    23.7\%    &    49.7\%    \\
Jets          &    11.6\%    &    31.3\%    \\
Cactus        &    8.1\%     &    10.6\%    \\
BlueSky       &    33.4\%    &    53.1\%    \\
Station       &    34.9\%    &    65.8\%    \\
SpinCalendar  &    24.3\%    &    62.4\%    \\
CatRobot      &    5.9\%     &    9.0\%     \\
RollerCoaster &    12.4\%    &    15.2\%    \\
\hline
\end{tabular}
\end{table}

\section{Conclusion}
\label{Sec::conclusions}
In this paper, an effective four-parameter affine motion compensation framework is proposed to better characterize the combination of rotation, zooming, and translation.
In the framework, a four-parameter affine motion model is firstly proposed and analyzed.
Then the four parameters are proposed to be coded in two manners: advanced affine motion vector prediction and affine model merge.
Especially, different from the traditional merge mode to regenerate a new affine motion model using the neighboring motion information, the affine model merge mode reuses the affine motion model of the neighboring blocks using affine mode.
Moreover, two fast motion compensation tools including one-step sub-pixel interpolation filter and affine interpolation-precision-based adaptive block size motion compensation are proposed to speed up the affine motion compensation process.
The proposed framework is implemented in the reference software of High Efficiency Video Coding (HEVC).
The experimental results show that the proposed affine motion compensation framework can achieve much better rate distortion performances compared with the HEVC anchor for the sequences with rich rotation or zooming motions.
The experimental results demonstrate the effectiveness of the proposed affine motion compensation framework.

In the current implementation, we only focus on a quite simple four-parameter affine motion model to characterize the combination of rotation, zooming, and translation.
However, how to effectively characterize other complex motions using high-order motion models remains an open issue.
Besides, the global high-order motion model is sometimes more effective than the local high-order motion model.
We will try to integrate the global and local high-order motion models into a whole framework in our future work.

\ifCLASSOPTIONcaptionsoff
  \newpage
\fi

% Generated by IEEEtran.bst, version: 1.12 (2007/01/11)

\end{document}